\documentclass[onecolumn,natbib,lscape]{aa}
\usepackage{natbib}
\usepackage{graphicx}
\usepackage{graphics}
\usepackage{longtable}
\usepackage{lscape}
\usepackage{amssymb}
\usepackage{txfonts}
\bibstyle{aa} 
\def\bb{} 
\def\arcsec{\ifmmode^{\prime\prime}\;\else$^{\prime\prime}\;$\fi}
\def\arcmin{\hbox{$^\prime$}}
\def\deg{\hbox{$^\circ$}}
\def\funits{\,{\rm erg ~cm^{-2} ~s^{-1}}}

\begin{document}
   \title{Chandra point-source counts in distant galaxy clusters}
   \author{M. Branchesi
           \inst{1,2}, 
           I. M. Gioia
           \inst{2}, 
           C. Fanti
           \inst{2,3}, 
           R. Fanti
           \inst{2,3},
           N. Cappelluti
           \inst{4}}

   \offprints{M. Branchesi}

   \institute{Dipartimento di Astronomia, Universit\`a di Bologna,
              Via Ranzani 1, 40127 Bologna, Italy \\
                \email{m.branchesi@ira.inaf.it}
                           \and
            Istituto di Radioastronomia, INAF, Via Gobetti 101, 
            40129 Bologna, Italy \\
                \email{gioia@ira.inaf.it}
                          \and
           Dipartimento di Fisica, Universit\`a di Bologna, Via
           Irnerio 46, 40126 Bologna, Italy \\ 
                \email{rfanti@ira.inaf.it, cfanti@ira.inaf.it}
                          \and
           Max Planck Institute f\"ur Extraterrestrische Physik, \\
              Postfach 1312, 85741, Garching, Germany\\
               \email{cappelluti@mpe.mpg.de}}  

   \date{Received 7 August 2006; accepted 24 September 2006}

\abstract{With the superb angular resolution of the Chandra Observatory,
it is now possible to detect X--ray point sources, either embedded in galaxy 
clusters or along the cluster line of sight, which could not be resolved by 
previous instruments. This now allows studies of source counts in distant 
cluster fields.}
{We want to analyze the inner region of clusters of galaxies to check for 
the presence of any overdensity of X-ray point sources embedded 
in the gas diffuse emission. These point sources are possible AGN 
belonging to the clusters and could contaminate the cluster emission.}
{{\bb We used a sample of 18 distant ($ 0.25 <$~z~$< 1.01$) galaxy clusters 
from the Chandra archive to construct the $\log~N{-}\log~S$, in both the
soft and hard energy bands, for the X--ray point sources detected in the
central cluster region to be compared with the counts  of point sources 
detected in similarly deep fields without clusters.}}
{We find a $\sim$ 2 $\sigma$ excess of cluster region sources at the bright 
end of the $\log~N{-}\log~S$. The radial distribution of the brightest X--ray
point sources confirms this excess and indicates that it is confined to the 
inner 0.5~Mpc of the cluster region.}
{The results suggest the possible existence of X--ray sources belonging to the
cluster (most probably AGN, given their 0.5--10 keV luminosity ranging from 
10$^{43}$ to 10$^{44}$ erg s$^{-1}$): on average one every three clusters. 
Unlike previous studies, which have mainly investigated the point-source 
population in the vicinity of the galaxy clusters, the present study analyzes 
the content of point sources within the 1 Mpc region covered by the cluster 
extent. Our work confirms the findings of other investigators who analyzed the 
central 1 Mpc region of more massive clusters and/or groups in a similar 
redshift range. The X--ray source excess found here is much smaller than 
the excess of radio galaxies found recently in high-z X--ray selected 
clusters, possibly due to the better sensitivity of the radio observations.}
   

\authorrunning{Branchesi et al.}
\titlerunning{X--ray point source counts in areas covered by distant clusters}

\maketitle

\section{Introduction}
Clusters of galaxies harbor a wide diversity of galaxy populations, so they 
are ideal laboratories for studying galaxy formation and evolution.
While there is clear evidence of the evolution in cluster galaxies 
\citep{Bu78,Bu84}, the evolution with redshift of cluster active galactic 
nuclei (AGN), as well as the possible prevalence of AGN in cluster 
environment with respect to the field, is still an unresolved issue. 
A related open debate is whether the cluster environment plays a role in 
the probability of galaxies to develop star-forming or AGN activity. 

\smallskip
In the radio domain, searches for active galaxies in nearby galaxy clusters 
have been carried out for a long time
\citep[see among others][]{Ow75,fan84,lo95a,led95b}.
The Radio Luminosity Function (RLF) of nearby cluster radio galaxies
was found to be  statistically indistinguishable from that of the field,
both in shape and  normalization \citep{lo96}. Recently, however,
a radio survey of high redshift galaxy clusters \citep{Br06} has provided
evidence of changes in the RLF of the distant cluster radio galaxies as
compared to the local one. All these studies show  that radio galaxies in
clusters are strongly centrally concentrated and that their radial
distribution essentially follows that of early type galaxies. As a result
of the shape of the RLF and of the high sensitivity of radio telescopes, the
typical radio luminosities found are not very high. Therefore the nuclear
activity of these radio galaxies is not prominent at optical and/or
X--ray wavelengths. On the other hand, studies from optical surveys seem to
suggest that AGN are relatively rare in cluster environments. 
Cluster members that show  evidence of AGN activity in their optical 
spectra are only 1\% of all cluster galaxies, while AGN are more common 
(5\%) in the field  population  \citep{Dr85}. Furthermore, there is no 
evidence of any  increase with redshift of the AGN fraction in clusters up to 
z~$\sim$~5 \citep{Dr99}.

\smallskip 
Since the pioneering work of \cite{He91}, who used
{\it ROSAT} data to first reveal the presence of significantly more
X--ray point sources around Abell 2256 (z=0.06) than one would expect
by chance, several studies have appeared in the literature that indicate
X--ray point-source excesses in the vicinity of low- and high-z
clusters with respect to the field. Many of these sources have been
confirmed to be cluster members in several cases. At low redshift the
excess seems to be largely due to the low-luminosity AGN ($\sim$10$^{41}$
erg s$^{-1}$) associated with the cluster \citep{La98,Su02}.  At
higher redshift, more luminous ($\sim$10$^{42-43}$ erg s$^{-1}$) AGN
are found in the fields of several clusters. It is evident from these 
studies that there is a population of obscured, or at least
optically unremarkable, AGN in galaxy clusters. An AGN
identification may not be obvious at optical wavelengths where the
dusty AGN or those with weak emission lines may be not recognized as
such \citep{Ma02}. {\bb A more recent work by \cite{Ma06}, who completed a
spectroscopic survey of X--ray point sources in eight low-redshift clusters
(z$<$0.3), finds that cluster galaxies host AGN more frequently than
previously thought, a factor 5 higher than found by \cite{Dr85}.}  Thus the
optical spectroscopic surveys alone can underestimate the number of
AGN in clusters. This is one of the reasons for the recently increased
popularity of X--ray and radio wavebands for AGN detection at high redshift.

\smallskip
The capabilities of the current generation of X--ray observatories
like \textit{Chandra} \citep{Van97} or XMM-Newton \citep{Sr01,Tur01}
have triggered and extended to higher redshift these type of analyses
\citep[among others][]{Ca01,Pen02,Mol02,Ca05} thereby enabling studies 
with very fine spatial details. The superb angular resolution of the {\em
Chandra} observatory, as well as the high sensitivity over the full
X--ray band, now allow the detection of those X--ray point sources
within the cluster region, which could not be resolved from the
diffuse emission with previous instruments. The works mentioned above
have shown an abundance of point sources in the direction of
individual high-z clusters as compared to the field. Different
hypotheses for the apparent overdensities are examined by the
different authors: 1) statistical variance of cosmic background
sources, 2) gravitationally lensed background sources, 3) AGN/quasars
and/or powerful starburst galaxies associated with the clusters. The
last hypothesis is now considered the most plausible one. 

\smallskip
\cite{Ca05} performed the first systematic search for X--ray point sources 
at the outskirts of 10 {\em Chandra} high-z clusters (0.24 $<$ z $<$ 1.2) 
and found a factor 2 overdensity $> 2 \sigma$ significance level in 40\%
of clusters fields. They speculate that the most likely astrophysical
interpretation of the overdensity is that the X--ray sources are AGN
that trace the filaments connected to the clusters.  This speculation
is supported by the results of \cite{Ca07}, who show that 
AGN and clusters are strongly correlated on scales between 2.5 and 
10 Mpc (smaller separations were not investigated). Further support for 
this interpretation comes from \cite{De04}, who studied the field 
surrounding the cluster 3C\,295, which clearly exhibits a strong and 
asymmetric clustering of X--ray sources on scales of a few arcminutes.

\smallskip
Most of the above-mentioned studies analyzed the
whole field around the clusters. In recent years it has become
evident the importance of the location of AGN with respect to the
cluster center, which could constrain different scenarios of AGN
triggering. The study by \cite{Jo03} of MS\,1054-03 at z $=$ 0.83, for
instance, indicates that the X--ray AGN excess avoids the central
regions of the cluster. Interestingly, AGN are not distributed
randomly within the cluster but tend to populate the outer 1--2 Mpc,
suggesting that AGN activity is triggered by recent infall at the
cluster outskirt.  In a recent work, \cite{Ru05} study the spatial
distribution of a sample of 508 X--ray point sources detected in the
soft 0.5--2.0 keV band in \textit{Chandra} observations of 51 clusters
(0.3$<$z$<$0.7) belonging to the MAssive Cluster Survey \citep[MACS;
][]{Eb01}.  The surface density of the X--ray point sources computed
in the cluster rest frame exhibits an 8 $\sigma$ excess within 3.5 Mpc 
of the cluster centers. The authors resolve two distinct components of the
excess, namely a central excess of AGN {\bb (at $\simeq$ 4 $\sigma$
within 0.5 Mpc, as can be seen from their Fig. 2)} and a broad secondary 
excess observed at about the virial radius, separated by a depletion 
region around 1.5 Mpc.  They suggest that the central excess may be due to
galaxy mergers and tidal interactions involving the central giant elliptical 
galaxies. The second excess could be caused by increased AGN activity at 
the cluster-field interface due to merger-induced accretion onto massive 
black holes. 

\smallskip
In conclusion it seems that X--ray selected AGN
are broadly distributed across the cluster.  This could be a
consequence of the different processes occurring close to the center
of the cluster with respect to those occurring at the cluster-field
interface. As mentioned earlier, radio galaxies are instead very
centrally peaked.  These two different distributions (X--ray AGN vs
radio galaxies) are not necessarily at variance since at the typical
values of the radio luminosity \citep[see for instance][]{Fa04} the
X--ray emission drops below the current X--ray detection levels.

\smallskip
In order to explore and investigate further
whether and how the cluster environment plays a role in the statistical
AGN properties of galaxies, the present paper focuses on the properties
of point sources detected in the inner region of high-z clusters using
the X--ray energy band.  Our X--ray analysis is limited to the central
$\le$ 1 Mpc cluster region, so as to detect point sources embedded
in the diffuse emission of the cluster gas. Since the number of sources
expected in this area is not statistically significant for single
cluster studies, we used a combined sample of eighteen clusters
observed by \textit{Chandra}. The redshift range is chosen to cover
moderate to high redshift clusters ($ 0.25 <$~z~$< 1.01$) in order to
look for a possible redshift dependence of any excess found.  All
errors in this paper are at the 1 $\sigma$ confidence level, unless
otherwise noted.  Throughout the paper, we use a $\Lambda$CDM
cosmology \citep{Sp03} with $\rm H_{0} = 70 ~km ~s^{-1} ~Mpc^{-1}$ and
$\Omega_{\rm m} = 1 -\Omega_{\Lambda} = 0.3$.

\section{The cluster sample}

The sample used in this study consists of 18 clusters of galaxies observed 
by the {\em Chandra} observatory with  redshift in the range z = 0.25--1.01. 
The choice of the redshift interval (z $>$ 0.25 and up to z $\sim$ 1) was 
dictated by the requirement to select clusters that are reasonably matched 
in size by the field of view of the \textit{Chandra} CCDs and, at the same
time, at a cosmologically significant distance. We retrieved only
those observations from the \textit{Chandra} archive with exposure times 
greater than 30 ks to be able to detect low surface-brightness 
clusters and the faint point sources projected against them. 
{\bb The choice of the exposure time was also dictated by the requirement of
having enough count statistics to accurately measure the cluster 
temperatures that we will present in a future paper (Branchesi et al. in
preparation) where the effect of the X--ray point sources on
the cluster properties will be addressed}.
Table~\ref{tab1} lists the sample parameters and details of the 
{\em Chandra} observations. The columns contain the following information:
 \begin{itemize}
 \item [--] Column 1: Cluster name
 \item [--] Column 2: Spectroscopic redshift tabulated in the literature
 \item [--] Column 3-4: Right ascension and declination 
            (Equatorial J2000, HH MM SS.S, +DD MM SS.S) of the
            centroid of the {\em Chandra} photon distribution 
            in the 0.5--5 keV energy band assumed as the cluster center  
 \item [--] Column 5: Identification number of the observation
 \item [--] Column 6: Detector where the aimpoint lies (I, for  
           {\em ACIS-I} or S, for {\em ACIS-S}) 
 \item [--] Column 7: Observation mode (F for FAINT or VF for VFAINT)
 \item [--] Column 8: Exposure time in ks corresponding to the nominal
            exposure filtered to exclude time periods of high background
 \item [--] Column 9: Column density of Galactic hydrogen in units of
            10$^{20}$ cm$^{-2}$, obtained from the {\em Chandra} 
            X--ray Center (CXC) Proposal Planning Tool Colden (Galactic 
            Neutral Hydrogen Density Calculator):  NRAO-compilation by 
            \cite {Di90}.  
 \item [--] {\bb Column 10: Estimate of the luminosity limit for a cluster 
            point source in erg s$^{-1}$ in the 0.5--2.0 keV energy band. 
            This limit is computed using the flux corresponding to 90\% of 
            the sky coverage of each cluster (see Appendix~\ref{appendix:Sec3}
            for details)
 \item [--] Column 11: Same as Column 10 but for the 2.0--10.0 keV
            energy band }
\end{itemize}

\tabcolsep 0.1cm
\begin{table}
\begin{center}
\caption{Cluster sample parameters and details of Chandra observations}
\scriptsize
\small
\begin{tabular}{lcccrccrccc}
\hline 
\hline
Cluster name    &      z & RA     &      DEC  &Obs.ID& ACIS & Mode
& Exp. & ${\rm N_H}$& L$^{lim}_{0.5-2.0}$& L$^{lim}_{2.0-10.0}$\\ 
        &        & $hh ~mm ~ss$ & ~~\deg ~~~~\arcmin ~~~~\arcsec &   &  & & ks &
\small{$10^{20}$ cm$^{-2}$} & \small{$10^{42}$ cgs} & \small{$10^{42}$ cgs} \\ 
(1) & (2) & (3) & (4) & (5) & (6) & (7) & (8) & (9) & (10) & (11)\\
\hline
\hline
Abell\,2125           & 0.246& 15 41 12&   $+$66 16 01&  2207& I & VF & 79.7 & 2.77 & 0.13 & 0.56 \\
ZW\,CL\,1454.8$+$2233 & 0.258& 14 57 15&   $+$22 20 33&  4192& I & VF & 91.4 & 3.22 & 0.23 & 0.74 \\
MS\,1008.1$-$1224     & 0.302& 10 10 32&   $-$12 39 23&  926 & I & VF & 44.2 & 6.74 & 0.44 & 1.57 \\
ZW\,CL\,0024.0$+$1652 & 0.394& 00 26 35&   $+$17 09 39&  929 & S & VF & 36.7 & 4.19 & 0.34 & 2.22 \\
MS\,1621.5$+$2640     & 0.426& 16 23 36&   $+$26 34 21&  546 & I & F  & 30.0 & 3.59 & 0.81 & 3.41 \\  
RXJ\,1701.3$+$6414    & 0.453& 17 01 24&   $+$64 14 10&  547 & I & VF & 49.5 & 2.59 & 0.64 & 2.67 \\
CL\,1641$+$4001       & 0.464& 16 41 53&   $+$40 01 46&  3575& I & VF & 44.0 & 1.02 & 0.67 & 2.62 \\
V\,1524.6$+$0957      & 0.516& 15 24 40&   $+$09 57 48&  1664& I & VF & 49.9 & 2.92 & 0.89 & 3.29 \\
MS\,0451.6$-$0305     & 0.539& 04 54 12&   $-$03 00 53&  902 & S & F  & 41.5 & 5.18 & 0.73 & 4.12 \\
V\,1121$+$2327        & 0.562& 11 20 57&   $+$23 26 27&  1660& I & VF & 66.9 & 1.30 & 0.73 & 3.00 \\
MS\,2053.7$-$0449     & 0.583& 20 56 21&   $-$04 37 51&  1667& I & VF & 43.5 & 4.96 & 1.32 & 4.91 \\ 
V\,1221$+$4918        & 0.700& 12 21 26&   $+$49 18 30&  1662& I & VF & 79.4 & 1.44 & 1.18 & 4.62 \\
MS\,1137.5$+$6625     & 0.782& 11 40 22&   $+$66 08 18&  536 & I & VF & 117.5& 1.18 & 0.81 & 4.04 \\ 
RDCSJ\,1317$+$2911    & 0.805& 13 17 21&   $+$29 11 19&  2228& I & VF & 111.3& 1.04 & 0.85 & 3.59 \\  
RDCSJ\,1350$+$6007    & 0.805& 13 50 48&   $+$60 06 54&  2229& I & VF & 58.3 & 1.76 & 1.77 & 7.26 \\  
RXJ\,1716.4$+$6708    & 0.813& 17 16 49&   $+$67 08 26&  548 & I & F  & 51.5 & 3.71 & 2.17 & 9.45 \\
MS\,1054.4$-$0321     & 0.830& 10 56 59&   $-$03 37 37&  512 & S & F  & 67.5 & 3.67 & 1.07 & 6.61 \\  
WARPJ\,1415.1$+$3612  & 1.013& 14 15 11&   $+$36 12 00&  4163& I & VF & 89.2 & 1.10 & 1.93 & 7.54 \\
\hline
\label{tab1}
\end{tabular}
\normalsize
\end{center}
\end{table}

\medskip\noindent
With the exception of six clusters coming from the {\em Einstein} Medium 
Sensitivity Survey \citep[EMSS;][]{Gi90}, all the clusters were originally  
discovered  in {\em ROSAT} surveys, either the {\em ROSAT} All Sky Survey 
\citep{vog99} or serendipitous surveys from pointed observations. Four 
clusters come from the 160 Square Degrees {\em ROSAT} Survey 
\citep[][]{Vi98,Mu03}, three from the {\em ROSAT} Deep Cluster Survey 
\citep[RDCS;][]{Ro98}, one from the Wide Angle {\em ROSAT} Pointed Survey 
\citep[WARPS;][]{Pe02}, and one, RXJ\,1716$+$6708 \citep{Gi99}, is part of 
the NEP survey \citep{Gio03,He06}. Three clusters, ZW\,CL\,0024.0+1652, 
ZW\,CL\,1454.8+2233,  Abell\,2125 are instead optically selected clusters.

\section{Source detection strategy}
\label{redu}
Data reduction was performed using version 3.2.1 of the CIAO software
({\it Chandra} Interactive Analysis of Observations; see web page 
{\em http://cxc.harvard.edu/ciao/index.html)}. The details are given
in Appendix~\ref{appendix:Sec1}.

\smallskip\noindent
For each {\em ACIS} CCD chip, two separate images were extracted from the 
event file for source detection at the raw resolution of 0.492 arcsec 
pixel$^{-1}$. The two images are characterized by the following energies:
a soft-energy image (0.5--2.0 keV) and a hard-energy image (2.0--7.0 keV). 
The cut-off below 0.5 keV is necessary due to the steep drop off of the 
quantum efficiency and to the steep rise observed in the background rate 
due to charged particles. The cut-off above 7.0 keV is necessary due to
the decrease in the effective area of \textit{Chandra} and to the increase 
in the instrumental background, which limits the detection efficiency of
sky and source photons. 

\smallskip\noindent
Sources were detected using the WAVDETECT algorithm 
\citep{Fre02}, included in the CIAO software package.
The significance threshold used for source detection was set to
the inverse of the total number of pixels, e.g. $\sim 10^{-6}$ for a 
1024 $\times$ 1024 pixels field. This is equivalent to stating that 
the expected number of false sources is one over the area of a single 
full-resolution {\em ACIS} chip.
Wavelet scales were chosen in nine steps of $(\sqrt{2})^{i}$
pixel $(i=0,..,8)$ starting from 0.492\arcsec, i.e. (0.5\arcsec-- 8.0\arcsec) 
to cover a wide range of source sizes, accommodating extended sources and 
the variation in the PSF as a function of the off-axis angular distance, 
$\Theta$ (i.e. the distance of the source from the aimpoint).

\smallskip\noindent
The algorithm also uses an exposure map for each energy band to account for
variations in the effective exposure across the {\em ACIS} field of view. 
To consider the photon-energy dependence of the effective exposure time,  
exposure maps were created at a single energy resolution representative of 
the mean energy of the photons in each band: 1.0 keV for the soft band and 
4.0 keV for the hard band.

\section{X--ray point-source sample}
\label{sample}
Even if WAVDETECT is a detection algorithm, it supplies a list of source 
parameters that is very useful for photometric analysis. After running 
the algorithm on the different energy images, these parameters were used 
to build a sample of point sources.

\smallskip\noindent
Following \cite{Ma03} and \cite{Jo03}, the source lists were built by 
accepting sources with a signal-to-noise ratio ($S/N$) greater than 3.0.
This limit is a reasonably conservative one that guarantees
the reliability of the sample sources. 
The source significance is defined as
\begin{eqnarray}
S/N = C/(1 + \sqrt{0.75+B})
\label{eqn1}
\end{eqnarray} 
where $C$ are the net source counts, and $B$ are the background counts
within the ``source cell", a region defined by WAVDETECT, which is
assumed to contain all the source counts \citep{Fre02}. The
denominator of Eq.~\ref{eqn1} is an approximate expression for the
error on the background counts from \cite{Gh86}, who gives the upper
confidence level equivalent to 1 $\sigma$ Gaussian error for small
number statistics.  The definition of source significance is
computationally convenient for defining a flux limit (see
Appendix~\ref{appendix:Sec3}) in each energy band. For a number of
clusters we noticed that the detection algorithm tends to consider as
point sources some slightly extended emission regions close to the
cluster center. In most cases, these sources are simply cluster clumps
of the thermal gas rather than central X--ray point sources.  A visual
inspection of all the detected sources enabled us to tentatively
eliminate these dubious point source identifications from the sample.

\smallskip\noindent 
A total of 119 X--ray sources were detected in the {\bb searched 
cluster area}. Of these, 41 sources were detected only in
the soft band, 24 only in the hard band, and 54 are common to both
bands. The search radius $\rm R_{ext}$ (listed in Table~\ref{tab2} for
each cluster) is the radius that includes the cluster region where
diffuse emission is still detectable. It is namely the radius at which
the cluster surface-brightness profile merges into the background, and
beyond which no further significant cluster emission is detectable.
The total area covered by the clusters is $\sim$ 0.083 deg$^2$.  The
survey is complete down to a flux limit of $2.7 \times 10^{-15}$
$\funits$ in the soft energy band and to $0.8 \times 10^{-14}$
$\funits$ in the hard energy band, corresponding to 100\% of the
respective sky coverages (see Appendix~\ref{appendix:Sec3}). {\bb
However there are a number of individual clusters surveyed to a fainter
point source flux}.

\smallskip\noindent 
The source list is given in Table~\ref{tab2}. A
detailed description of how the X--ray source parameters are computed
is given in Appendix~\ref{appendix:Sec2}.  The columns of
Table~\ref{tab2} contain the following information:
\begin{itemize}
\item [--] Column 1: Cluster name
\item [--] Column 2: Search radius, $\rm R_{ext}$, in arcsec 
\item [--] Column 3: Source identification number
\item [--] Column 4-5: WAVDETECT source position; Right Ascension and 
Declination  (Equatorial J2000, HH MM SS.S, +DD MM SS.S)
\item [--] Column 6-7: Net counts in the soft (0.5--2.0 keV) and hard 
 (2.0--7.0 keV) energy bands. Asterisks indicate sources not used in 
Sect.~\ref{cap4} for the computation  of the source counts since their 
flux is smaller than the flux corresponding to 20\% of the sky coverage 
(see Appendix~\ref{appendix:Sec3})
\item [--] Column 8: Observed X--ray flux in units of 10$^{-15}$ $\funits$ 
in the soft (0.5--2.0 keV) band. The flux and the associated error have been 
calculated as described in Appendix~\ref{appendix:Sec2}. 
\item [--] Column 9: Correction factor to be applied to the observed 
           soft X--ray flux to obtain the Galactic absorption corrected 
           X--ray flux.
\item [--] Column 10: Same as Column 8 but for the  hard (2.0--10.0 keV) band.
\end{itemize}

\tabcolsep 0.1cm
\begin{longtable}{llcccrrrrrr}
\caption{X--ray cluster region point sources}
\\
\hline 
\hline
Cluster name& $\rm R_{ext}$ &\# & RA & DEC & C$_{0.5-2.0}$ & C$_{2.0-7.0}$ & S$_{0.5-2.0}$  & c$_{N_{H}}$ & S$_{2.0-10.0}$\\ 
            & \arcsec  &   &$~~hh ~mm ~ss.s$ & ~~\deg ~~~~\arcmin ~~~~\arcsec  &     &  &  \small{$10^{-15}$ cgs} &     &  \small{$10^{-15}$ cgs}\\
(1) & (2) & (3) & (4) & (5) & (6) & (7) & (8) & (9) & (10)\\
\hline
\hline
\endhead
Abell\,2125       & 241 &  1& 15 40 39.4&   +66  17  13.2&  30.63  & 10.77&$  2.28\pm0.43$& 1.083&$  3.49\pm1.13$&\\
   	    	     &  &  2& 15 40 45.3&   +66  17  27.3&   9.77  &  7.75&$  0.73\pm0.24$&	 &$  2.52\pm0.96$&\\
   	    	     &  &  3& 15 40 46.7&   +66  13  21.0&  11.92  &	  &$  0.84\pm0.24$&	 &		 &\\
   	             &	&  4& 15 40 52.4&   +66  12  36.9& 104.72  & 39.61&$  7.28\pm0.74$&	 &$ 12.50\pm2.38$&\\
   	    	     &  &  5& 15 40 56.4&   +66  16  28.7& 420.99  &141.58&$ 30.96\pm1.76$&	 &$ 45.71\pm6.11$&\\
   	    	     &	&  6& 15 40 58.9&    +66  17  42.8&  13.82  &	  &$  1.05\pm0.30$&	 &		 &\\
   	    	     &	&  7& 15 41 00.4&    +66  19  03.0&  13.83  &	  &$  1.11\pm0.31$&	 &		 &\\
   	    	     &	&  8& 15 41 02.0&    +66  17  21.4&  82.37  & 27.88&$  6.25\pm0.71$&	 &$  9.17\pm2.01$&\\
   	    	     &  &  9& 15 41 02.0&    +66  16  27.2&  27.31  &	  &$  2.02\pm0.40$&	 &		 &\\
   	    	     &	& 10& 15 41 02.7&    +66  14  04.7&  36.60  & 15.70&$  2.62\pm0.44$&	 &$  5.01\pm1.38$&\\
   	    	     &	& 11& 15 41 09.2&    +66  14  49.0&   8.55  &	  &$  0.62\pm0.22$&	 &		 &\\
   	    	     &	& 12& 15 41 09.8&    +66  15  45.3&  10.37  &	  &$  0.77\pm0.26$&	 &		 &\\
   	    	     &  & 13& 15 41 12.4&    +66  17  17.1&  18.37  &  9.19&$  1.42\pm0.35$&	 &$ 3.06\pm1.10 $&\\
   	    	     &	& 14& 15 41 16.9&    +66  16  26.9&  10.14  &	  &$  0.78\pm0.27$&	 &		 &\\
   	    	     &	& 15& 15 41 17.4&    +66  19  24.0&  21.02  &	  &$  1.76\pm0.40$&	 &		 &\\
   	    	     &	& 16& 15 41 17.8&    +66  13  43.1&  19.59  & 12.72&$  1.45\pm0.33$&	 &$  4.18\pm1.26$&\\
   	    	     &	& 17& 15 41 26.2&    +66  13  41.4&	   &  6.73&		  &	 &$ 2.36\pm0.96 $&\\
   	    	     &	& 18& 15 41 27.3&    +66  17  41.5&  14.39  &	  &$  1.40\pm0.39$&	 &		 &\\
   	    	     &	& 19& 15 41 27.4&    +66  14  13.7&  19.64  & 16.55&$  1.77\pm0.41$&	 &$  6.51\pm1.76$&\\
   	    	     &	& 20& 15 41 28.3&    +66  12  47.5&   6.84  &	  &$  0.60\pm0.23$&	 &		 &\\ 
   	    	     &	& 21& 15 41 33.7&    +66  13  42.1&  16.58  &  7.45&$  1.30\pm0.32$&	 &$  2.54\pm1.00$&\\
   	    	     &	& 22& 15 41 37.3&    +66  15  07.1&   9.34  &	  &$  0.75\pm0.25$&	 &		 &\\   
   	    	     &	& 23& 15 41 41.1&    +66  16  42.0&   8.41  &	  &$  0.70\pm0.25$&	 &		 &\\
   	    	     &	& 24& 15 41 43.5&    +66  14  19.4&  15.76  &	  &$  1.28\pm0.34$&	 &		 &\\
   	    	     &  & 25& 15 41 45.0&    +66  15  10.7&	   &  9.61&		  &	 &$ 3.47\pm1.25$ &\\
\hline	    	     	 								 
ZW\,CL\,1454$+$2233 &200&  1& 14 57  9.7&    +22  23  04.0&  22.10  & 14.17&$  1.69\pm0.40$& 1.096&$  4.29\pm1.32$&\\
   	    	     &	&  2& 14 57 10.8&    +22  18  45.0&  46.94  & 21.56&$  3.13\pm0.47$&	 &$  6.11\pm1.47$&\\
   	    	     &	&  3& 14 57 12.2&    +22  21  42.4&  87.63  & 27.32&$  6.41\pm0.76$&	 &$  8.10\pm1.90$&\\
   	    	     &	&  4& 14 57 13.2&    +22  17  27.0&  25.65  & 15.59&$  1.71\pm0.34$&	 &$  4.43\pm1.23$&\\
   	    	     &	&  5& 14 57 14.8&    +22  19  33.5&  29.82  &	  &$  2.07\pm0.47$&	 &		 &\\
   	    	     &	&  6& 14 57 17.7&    +22  19  22.8&  35.03  & 17.83&$  2.58\pm0.48$&	 &$  5.46\pm1.48$&\\
   	    	     &	&  7& 14 57 21.0&    +22  23  35.3&  88.99  & 22.61&$  7.44\pm0.89$&	 &$  7.15\pm1.86$&\\
\hline		     	 											  
MS\,1008$-$1224   &172  &  1& 10 10 21.4&   -12  40  07.9&   5.90* &  7.84&$0.69\pm0.29$& 1.203&$  4.49\pm1.69$&\\  
   	    	     &	&  2& 10 10 24.7&   -12  40  16.9&	   & 13.74&		  &	 &$  8.81\pm2.57$&\\
   	    	     &  &  3& 10 10 26.4&   -12  38  10.9&   7.50  &	  &$  0.92\pm0.35$&	 &		 &\\
   	    	     &	&  4& 10 10 29.0&   -12  40  13.5&	   & 25.93&		  &	 &$ 15.39\pm3.47$&\\
   	    	     &	&  5& 10 10 32.3&   -12  39  34.8&  17.68  &      &$2.21\pm0.66$&   &     &\\
   	    	     &	&  6& 10 10 35.3&   -12  40  22.0&  47.40  & 16.31&$  6.45\pm0.97$&	 &$ 10.67\pm2.91$&\\ 
    	    	     &	&  7& 10 10 37.1&   -12  38  57.8&  16.82  &	  &$  2.47\pm0.66$&	 &		 &\\
        	     &	&  8& 10 10 39.4&   -12  41  09.2&   9.40  &	  &$  1.24\pm0.42$&	 &		 &\\
   	    	     &  &  9& 10 10 42.7&   -12  39  19.1&  20.88  &	  &$  2.84\pm0.65$&	 &		 &\\
   	    	     &  &  10& 10 10 41.8&   -12  40  02.3&  11.84  &	  &$  1.59\pm0.50$&	 &		 &\\
\hline	 	     	 											  
ZW\,CL\,0024$+$1652&118 &  1& 00 26 31.0&   +17  10  30.4& 15.71   &	  &$  1.34\pm0.34$& 1.125&		 &\\
   	    	     &  &  2& 00 26 31.1&   +17  10  17.3&235.11   & 66.39&$ 20.81\pm1.38$&	 &$ 43.79\pm7.25$&\\
   	    	     &  &  3& 00 26 31.7&   +17  10  22.6& 16.70   &	  &$  1.48\pm0.36$&	 &		 &\\
   	    	     &  &  4& 00 26 32.0&   +17  09  41.9&	   & 19.60&		  &	 &$ 12.52\pm3.17$&\\
   	    	     &  &  5& 00 26 33.0&   +17  07  59.9& 33.51   &  6.70&$  2.93\pm0.51$&	 &$  4.38\pm1.80$&\\
   	    	     &  &  6& 00 26 33.3&   +17  10  34.5&	   &  6.74&		  &	 &$  4.27\pm1.74$&\\
\hline		     	 											  
MS\,1621$+$2640     &148&  1& 16 23 29.0&    +26  34  46.8&   5.81* &	  &$  1.01\pm0.43$& 1.107&		 &\\
   	    	     &  &  2& 16 23 33.9&    +26  35  24.9&   7.65  &  7.76&$  1.32\pm0.49$&	 &$  6.45\pm2.45$&\\
   	    	     &  &  3& 16 23 40.3&    +26  35  50.0&	   &  7.77&		  &	 &$  6.58\pm2.49$&\\ 
                     &  &  4& 16 23 43.7&    +26  32  44.7& 282.84  &107.61&$ 52.80\pm3.37$&	 &$94.81\pm13.50$&\\ 
                     &  &  5& 16 23 45.8&    +26  33  35.2&	   & 15.34&		  &	 &$ 13.44\pm3.77$&\\ 
\hline		     
RXJ\,1701$+$6414   &108 &  1& 17 01 13.1&    +64  12  50.4&   5.88* &  7.61&$ 0.67\pm0.28$ &1.077 &$  3.94\pm1.52$&\\ 
                     &  &  2& 17 01 21.5&    +64  15  05.2&	   &  8.54&		  &	 &$  4.50\pm1.65$&\\ 
                     &  &  3& 17 01 28.3&    +64  13  32.7&	   & 32.31&		  &	 &$ 16.54\pm3.40$&\\ 
\hline		     	 										 
CL\,1641$+$4001    &89  &  1& 16 41 50.0&    +40  02  49.4&  13.82  &  7.79&$  1.99\pm0.54$& 1.030&$  4.57\pm1.72$&\\ 
   	    	     &  &  2& 16 41 50.3&    +40  01  45.7& 162.94  & 46.56&$ 22.74\pm1.95$&	 &$ 26.83\pm4.83$&\\ 
   	    	     &  &  3& 16 41 54.2&    +40  00  32.6& 172.48  & 93.56&$ 23.72\pm1.98$&	 &$ 53.58\pm7.85$&\\   
                     &  &  4& 16 41 53.6&    +40  01  45.0&   8.71  &	  &$  1.22\pm0.45$&	 &		 &\\
                     &  &  5& 16 41 55.4&    +40  01  43.0&	   &  7.59&		  &	 &$  4.38\pm1.70$&\\ 
                     &  &  6& 16 41 58.4&    +40  00  48.4&   9.89  &  6.89&$ 1.38\pm0.44$ &	 &$  3.98\pm1.58$&\\ 
\hline		     	 										  
V\,1524$+$0957    &148  &  1& 15 24 30.6&    +09  57  30.5&  12.77  & 13.76&$ 1.63\pm0.46$ &1.087 &$  7.20\pm2.09$&\\
                     &  &  2& 15 24 32.3&    +09  57  45.1&  23.74  & 18.80&$ 2.97\pm0.62$ &	 &$  9.75\pm2.48$&\\
                     &  &  3& 15 24 32.4&    +09  59  07.2&	   &  5.91&		  &	 &$  3.39\pm1.45$&\\
                     &  &  4& 15 24 35.5&    +09  58  22.3&   7.82  &  9.74&$ 0.99\pm0.36$ &	 &$  5.24\pm1.79$&\\
                     &  &  5& 15 24 38.0&    +09  58  53.2&  21.56  & 10.81&$ 2.55\pm0.56$ &	 &$  5.45\pm1.77$&\\
                     &  &  6& 15 24 42.4&    +10  00  01.0&   6.89  &  6.83&$ 0.85\pm0.33$ &	 &$  3.69\pm1.48$&\\
                     &  &  7& 15 24 43.4&    +09  55  36.0& 221.31  & 95.38&$28.86\pm0.22$ &	 &$ 50.83\pm7.45$&\\
                     &  &  8& 15 24 43.7&    +09  56  05.1&   7.55  &	  &$ 0.96\pm0.36$ &	 &$		$&\\
\hline		     	 							       
MS\,0451.6$-$0305 &148  &  1& 04 54 12.9&   -03  00  46.8& 42.92   &	  &$3.33\pm0.58 $ &1.156 &$		$&\\
                     &  &  2& 04 54 10.9&   -03  01  24.3& 13.26   &	  &$1.13\pm0.35 $ &	 &$		$&\\
                     &  &  3& 04 54 12.3&   -02  59  11.3&  6.74   &  7.78&$0.54\pm0.21 $ &	 &$  4.47\pm1.70$&\\ 
                     &  &  4& 04 54 16.0&   -03  02  32.2&	   &  6.76&		  &	 &$  4.01\pm1.64$&\\	
\hline		     	 											  
V\,1121$+$2327    &128  &  1& 11 20 49.7&    +23  27  21.2&  54.77  & 26.67&$ 4.70\pm0.65$ &1.039 &$ 10.10\pm2.23$&\\ 
                     &  &  2& 11 20 49.8&    +23  26  30.4&   8.83  & 13.83&$ 0.75\pm0.25$ &	 &$ 5.17\pm1.50 $&\\
                     &  &  3& 11 20 54.0&    +23  27  04.9& 160.70  & 44.51&$13.98\pm1.19$ &	 &$ 16.89\pm3.09$&\\   
                     &  &  4& 11 20 58.8&    +23  26  29.6&	   &  8.48&		  &	 &$ 3.25\pm1.20 $&\\
                     &  &  5& 11 21 04.7&    +23  25  11.4&	   &  6.78&		  &	 &$ 2.64\pm1.07 $&\\
\hline		     
MS\,2053$-$0449   &118  &  1& 20 56 14.3&   -04  37  16.8&   9.81  &	  &$ 1.36\pm0.44$ &1.149 &		 &\\
                     &  &  2& 20 56 18.7&   -04  39  14.6&   7.91  &	  &$ 1.03\pm0.37$ &	 &		 &\\
\hline		    	 											  
V\,1221$+$4918    &143  &  1& 12 21 12.6&    +49  19  19.1&   6.68  &	  &$ 0.52\pm0.21$ &1.043 &		 &\\
                     &  &  2& 12 21 18.1&    +49  16  35.6&	   &  7.82&		  &	 &$ 2.47\pm0.93 $&\\
                     &  &  3& 12 21 20.1&    +49  18  44.0&  40.51  &153.94&$31.03\pm1.83$ &	 &$50.08\pm6.59 $&\\
                     &  &  4& 12 21 26.3&    +49  18  04.1&	   & 12.18&		  &	 &$ 3.92\pm1.23 $&\\
                     &  &  5& 12 21 30.9&    +49  17  57.3&   9.29  &	  &$ 0.76\pm0.26$ &	 &		 &\\
                     &  &  6& 12 21 29.1&    +49  16  43.4&   7.82  &  8.72&$ 0.57\pm0.21$ &	 &$ 2.76\pm9.92 $&\\
\hline		     	 											  
MS\,1137$+$6625   &103  &  1& 11 40 06.2&    +66  08  18.2&   6.99  &	  &$ 0.30\pm0.12$ &1.035 &		 &\\
                     &  &  2& 11 40 12.8&    +66  07  33.0&   9.64  & 21.33&$ 0.41\pm0.14$ &	 &$ 4.70\pm1.15 $&\\
                     &  &  3& 11 40 20.4&    +66  07  30.5&  39.55  & 22.19&$ 1.67\pm0.27$ &	 &$ 4.87\pm1.18 $&\\
                     &  &  4& 11 40 31.2&    +66  08  58.2& 860.74  &263.18&$36.44\pm1.43$ &	 &$59.46\pm7.50 $&\\
                     &  &  5& 11 40 33.7&    +66  07  39.6&   7.71  & 19.63&$ 0.32\pm1.18$ &	 &$ 4.35\pm1.10 $&\\
\hline		     	 											  
RDCSJ\,1317$+$2911 &69  &  1& 13 17 18.9&    +29  11  11.1&  62.43  & 24.59&$ 3.22\pm0.42$ &1.031 &$ 5.54\pm1.27 $&\\
                     &  &  2& 13 17 20.7&    +29  12  01.6&   6.85  &	  &$ 0.35\pm0.15$ &	 &		 &\\
                     &  &  3& 13 17 22.0&    +29  11  24.2&	   & 21.58&		  &	 &$ 4.89\pm1.18 $&\\
                     &  &  4& 13 17 23.5&    +29  11  49.5&   7.70  &	  &$ 0.41\pm0.15$ &	 &		 &\\
\hline		     	 											  
RDCSJ\,1350$+$6007&128  &  1& 13 50 37.8&    +60  08  21.2&  39.73  & 13.78&$ 4.15\pm0.67$ &1.052 &$ 6.11\pm1.78 $&\\
                     &  &  2& 13 50 39.8&    +60  05  06.8&	   &  6.85&		  &	 &$ 2.95\pm1.18 $&\\
                     &  &  3& 13 50 43.0&    +60  06  09.3&   6.92  &	  &$ 0.68\pm0.26$ &	 &		 &\\
                     &  &  4& 13 50 46.1&    +60  06  58.2&	   & 11.65&		  &	 &$ 5.03\pm1.58 $&\\
                     &  &  5& 13 50 50.2&    +60  08  01.5&   7.86  &	  &$ 0.98\pm0.35$ &	 &		 &\\
                     &  &  6& 13 50 50.4&    +60  06  20.5&   6.72  &	  &$ 0.67\pm0.26$ &	 &		 &\\
                     &  &  7& 13 50 57.0&    +60  07  28.6&	   &  9.79&		  &	 &$ 4.59\pm1.56 $&\\
                     &  &  8& 13 50 57.7&    +60  08  13.7& 123.40  & 21.74&$13.04\pm1.25$ &	 &$ 9.63\pm2.31 $&\\
                     &  &  9& 13 51 04.6&    +60  06  27.5&  11.89  & 10.86&$ 1.22\pm0.37$ &	 &$ 4.76\pm1.54 $&\\
\hline		     	 											  
RXJ\,1716$+$6708   &108 &  1& 17 16 36.9&    +67  08  30.0&   8.71  & 55.42&$ 1.05\pm0.36$ &1.111 &$ 32.1\pm5.48 $&\\
                     &  &  2& 17 16 37.6&    +67  07  31.0&   8.85  &	  &$ 1.02\pm0.34$ &	 &		 &\\
                     &  &  3& 17 16 42.2&    +67  06  59.8&  17.33  &	  &$ 1.84\pm0.45$ &	 &		 &\\
                     &  &  4& 17 16 51.7&    +67  08  54.8&  55.57  & 54.10&$ 5.63\pm0.77$ &	 &$ 26.71\pm4.60$&\\  
                     &  &  5& 17 16 53.1&    +67  07  50.2&  46.09  & 16.52&$ 4.75\pm0.71$ &	 &$  8.28\pm2.24$&\\
                     &  &  6& 17 17 07.4&    +67  08  40.0&	   & 5.85*&		  &	 &$  2.88\pm1.24$&\\
\hline		     	 											  
MS\,1054$-$0321   &128  &  1& 10 56 51.4&   -03  38  00.7&    7.77 &	  &$ 0.35\pm0.13$ &1.110 &		 &\\
                     &  &  2& 10 56 52.6&   -03  38  19.8&   23.59 &  7.75&$ 1.06\pm0.22$ &	 &$ 2.65\pm1.01 $&\\
                     &  &  3& 10 56 58.8&   -03  38  51.2&  272.10 &124.91&$12.29\pm0.77$ &	 &$ 42.08\pm6.08$&\\  
                     &  &  4& 10 57 04.9&   -03  38  21.2&	   &  9.71&		  &	 &$  3.31\pm1.14$&\\ 
\hline		      	 											  
WARPJ\,1415$+$3612 &79  &  1& 14 15 11.9&    +36  11  24.7&  16.59  & 10.72&$ 1.11\pm0.28$ &1.033&$ 3.03\pm0.99 $&\\
                     &  &  2& 14 15 12.4&    +36  13  03.9&   8.70  &	  &$ 0.60\pm0.21$ &	 &		 &\\
                     &  &  3& 14 15 13.5&    +36  12  10.4&	   & 22.51&		  &	 &$ 6.39\pm1.52 $&\\
                     &  &  4& 14 15 16.1&    +36  11  51.8&  40.76  & 19.70&$ 2.77\pm0.45$ &	 &$ 5.62\pm1.40 $&\\
\hline
\label{tab2}
\end{longtable}

\section{X--ray point-source number counts}
\label{cap4}
We used the source counts, or $\log~N{-}\log~S$ relationship, as a
statistical tool for investigating whether there is any point source
count excess in the regions of diffuse cluster emission with respect to
the fields void of visible clusters. Although integral source counts 
are usually given in the literature, we present here both the
differential and integral distributions. The differential number
counts are statistically more correct since source-count errors are
independent, differently from the integral number counts.  To compute
the $\log~N{-}\log~S$ relationship, we followed the method described in
\cite{Gi90}.  The source contributions are computed by properly
weighting each source with its visibility area, i.e. the area of sky
where the source intensity equals or exceeds the sensitivity limit
(see Appendix~\ref{appendix:Sec3} for details). The integral log (N
$>$ S) diagram is then built by summing, in decreasing flux order, the
contribution from each source. {\bb Since each cluster sky coverage 
rapidly decreases near the limiting flux for both the soft and hard bands, 
in order to prevent incompleteness effects,} we considered 
only those sources with a flux larger than the flux corresponding to 
20\% of the total sky area covered by each cluster. It is worth noting 
that, throughout the following sections, the soft X--ray fluxes are 
unabsorbed fluxes, i.e. corrected for the Galactic hydrogen column 
density absorption along the line of sight at each field position.

\subsection{X--ray source counts in the cluster regions}
\label{count1}

Following \cite{Cr70} and \cite{Mu73}, a Maximum Likelihood Method 
(MLM) that operates on the differential counts, was adopted to 
determine the slope of the number-flux distribution of the X--ray sources.
The first assumption is that the differential $\log~N{-}\log~S$ 
distribution may be described by a single power-law model of the form: 
\begin{eqnarray}
\frac{dN}{dS} = b\left( \frac{S}{S_{0}}\right)^{-\alpha1}.
\label{eqn2}
\end{eqnarray}
\noindent
The MLM  uses the 'unbinned' data that is very useful when one deals with 
a low number of sources. 

\smallskip\noindent
The MLM applied to the sources found in the regions covered by our
clusters (from now on {\it cluster region sources}) provides
the following results:
\begin{eqnarray}
\frac{dN}{dS} = 354 \pm 37 \left(\frac{S}{2 \times 10^{-15} {\rm ~erg~cm^{-2}~s^{-1}}}\right)^{-1.78 \pm 0.08} {\rm ~sources ~deg^{-2}}
\label{eqn3}
\end{eqnarray}
\noindent
for the 0.5--2.0 keV  band and
\begin{eqnarray}
\frac{dN}{dS} = 255 \pm 29 \left(\frac{S}{1 \times 10^{-14} {\rm ~erg~cm^{-2}~s^{-1}}}\right)^{-2.0 \pm 0.1} {\rm ~sources~~deg^{-2}}
\label{eqn4}
\end{eqnarray} 
for the 2.0--10.0 keV band.
 
\smallskip\noindent
The values for $b$ were calculated at the flux levels of 
$S_{0} = 2 \times 10^{-15}$~erg~cm$^{-2}$~s$^{-1}$ in the 
soft band and $S_{0} = 1 \times  10^{-14}$ erg cm$^{-2}$ s$^{-1}$ 
in the hard band. For the exposures considered here, these values
correspond to the centers of the sampled flux intervals.

\smallskip\noindent
The differential logarithmic source counts, in bins of  $\Delta\log~S=$ 0.2, 
are indicated in Fig.~\ref{fig1} by filled triangles.  The soft  (hard) 
$\log~N{-}\log~S$ for the X--ray point sources detected in the 
areas covered by the clusters is given in the top (bottom) panel.  
The counts and the errors are given by:
\begin{eqnarray}
dN =\sum_{i=1}^{n}\frac{1}{\Omega_{i}}~{\rm deg^{-2}}~~~~~~~ {\rm and}~~~~~~~~~~~~ 
\sigma_{dN} =\sqrt{\sum_{i=1}^{n}{(1/\Omega_{i}^2)}} ~{\rm deg^{-2}} 
\label{eqn5} 
\end{eqnarray}
where $n$ is the number of sources within each bin and $\Omega_{i}$ 
the visibility area of the i\textit{th} source. The thick solid lines 
represent the best fit obtained with the MLM described above. The 
dashed lines  indicate the 1 $\sigma$ uncertainties on the MLM fit 
parameters.

\medskip\noindent
To compare our results with those in the literature, we also 
calculated the cumulative source number counts N ($\ge$ S) and the 
corresponding errors as follows:

\begin{eqnarray}
N(>S)=\sum_{i=1}^{N_{tot}}\frac{1}{\Omega_{i}} ~~ {\rm deg}^{-2}~~~~~~~ {\rm and}~~~~~~~~~~~~
\sigma_{N(>S)} =\sqrt{\sum_{i=1}^{N_{tot}}{(1/\Omega_{i}^2)}}  ~~ {\rm deg}^{-2}
\label{eqn6}
\end{eqnarray} 

\noindent
where $N_{tot}$ is the total number of detected sources with a flux 
$\geq S$ and  $\Omega_{i}$  
the sky coverage corresponding to the flux of the i\textit{th} source.
The assumption made of a single power law converts Eq.~\ref{eqn2} into
\begin{eqnarray}
N(>\!S)=k\left(\frac{S}{S_{0}}\right)^{-\alpha}
\label{eqn7}
\end{eqnarray} 
for the integral distribution, where $\alpha = \alpha1-1$.  
The integral normalization $k$ is calculated by integrating  
Eq.~\ref{eqn2} between the flux limit and infinity:
\begin{eqnarray}
k = \frac{b}{\alpha1-1}~~. 
\label{eqn8} 
\end{eqnarray}
The errors on $k$ were calculated using the propagation of error
for $\alpha1$ and $b$. The results for the integral source counts
in the cluster fields are listed in Table~\ref{tab3}.

\smallskip\noindent
Figure~\ref{fig2} illustrates the cumulative source counts, indicated by
solid triangles, which were plotted using a step of $\Delta$\,log S $=$ 0.15.
The soft integral $\log~N{-}\log~S$ is shown in the top panel and the hard
integral $\log~N{-}\log~S$ in the bottom panel. Errors correspond 
to 1 $\sigma$ confidence level and are obtained using Eq.~\ref{eqn6}.
Note that the errors on the integral $\log~N{-}\log~S$ are not independent.

\subsection{X--ray source counts in the reference fields}
\label{count2}

The slope and/or normalization  of the source counts at a given
flux limit can provide information on the point-source population in the
cluster regions compared to the sources in fields without visible clusters
(from now on {\it field sources}). To check for the presence of such an 
effect, the results obtained in the area of the 18 clusters were compared 
with those obtained in four reference fields void of clusters with exposure 
times similar to the exposures of the cluster fields.

\smallskip\noindent
The reference fields used were the {\it Chandra} Deep
Field  South (CDFS), the Hubble Deep Field North (HDFN), the
Groth/Westphal strip area, and the Bootes field (see Table~\ref{tab3}).
In order to sample similar flux levels, the analysis was 
limited to observations with exposure on the order of 100ks.
The analysis was performed on the four {\em ACIS-I} CCD 
(16.8\arcmin$\times$16.8\arcmin). For the validity of our analysis 
and to avoid systematic errors, we ran exactly the same procedure
on the reference fields as was used to derive the $\log~N{-}\log~S$
of the cluster region  sources. The same source-detection routine, 
flux limit and flux estimate method have been adopted  (see 
Sects.~\ref{redu} and \ref{sample} and Appendix  for details).
The sky coverage of each reference field was constructed following the
method described in Appendix~\ref{appendix:Sec3}. 
The only difference is the binning of the background and exposure maps, 
which were chosen to be  $64 \times 64$ pixels  to reduce the computer time. 
The value of the flux limit of each bin is a representative mean 
of all the original pixels since the absence of a galaxy cluster in 
the field results in a much more uniform background.  The sky area of each 
field measures $\sim$ 0.078 deg$^{2}$. Following the same approach as the 
cluster region sources we considered only field sources with a flux greater 
than the flux corresponding to 20\% of the sky coverage of each field.

\medskip\noindent 
The best-fit results on the cumulative
$\log~N{-}\log~S$ in each reference field are summarized in
Table~\ref{tab3}.  Note that the literature reports source counts that 
are usually estimated using the observed fluxes, while we corrected for
the Galactic hydrogen column density to produce unabsorbed fluxes.
\tabcolsep 0.2cm
\begin{table}[htb]
\begin{center}
\caption{Integral source counts in clusters and in reference fields}
\scriptsize
\small
\begin{tabular}{lcccccccccc}
\hline 
\hline
Name & Obs.Id & $\tau$ & N$_H$ & N$_{0.5-2.0}$ &  N$_{2.0-7.0}$
& $\alpha_{0.5-2.0}$ & $\alpha_{2.0-10.0}$ & k$_{0.5-2.0}$ & k$_{2.0-10.0}$ \\ 
         &  &ks  & cm$^{-2}$& &   &    &   &  deg$^{-2}$ &  deg$^{-2}$ \\
(1) & (2) & (3) & (4) & (5) & (6) & (7) & (8) & (9) & (10)\\
\hline
\hline
\\
Clusters &     &    &  &  92 & 77&0.78~$\pm$~0.08 & 1.00~$\pm$~0.10 & 454~$\pm$~66 & 255~$\pm$~39\\
\hline
\\
Bootes   & 3130 & 107 & 1.15 & 90 & 84 &0.76~$\pm$~0.06&1.11~$\pm$~0.10 & 410~$\pm$~54 & 217~$\pm$~31\\
HDFN     & 3389 & 109 & 1.50 & 96 & 63 &0.78~$\pm$~0.08&0.90~$\pm$~0.10  & 426~$\pm$~62 & 240~$\pm$~41\\  
CDFS     &  582 & 128 & 0.80 &109 & 81 &0.80~$\pm$~0.06&1.03~$\pm$~0.10  & 396~$\pm$~48 & 210~$\pm$~31\\
Groth    & 4357 &  85 & 1.30 & 83 & 52 &0.80~$\pm$~0.08&0.94~$\pm$~0.12  & 432~$\pm$~63 & 217~$\pm$~41\\
\\
Unified  &     &     &   &  378 & 280 &0.79~$\pm$~0.04 & 1.01~$\pm$~0.04 & 416~$\pm$~30 & 219~$\pm$~16\\
\hline
\\
\label{tab3}
\end{tabular}
\end{center}
\vspace{-1cm}
\begin{itemize}
\item [--] Column 1: Reference field name
\item [--] Column 2: Identification number of the observation
\item [--] Column 3: Total net exposure time after 'flare' cleaning
\item [--] Column 4: Hydrogen Galactic column density value, N$_H$, 
       in units of 10$^{20}$ cm$^{-2}$. 
\item [--] Column 5-6: Number of sources detected in the soft 
         and hard bands with flux larger than the flux corresponding 
         to 20\% of the sky coverage  
\item [--] Column 7-8: The integral $\log~N{-}\log~S$ power law slope 
          and 1 $\sigma$ confidence level error for the soft and
          hard bands
 \item [--] Column 9-10: The integral $\log~N{-}\log~S$ normalization 
        and 1 $\sigma$ confidence level error calculated in the soft band 
        at a flux of $2 \times 10^{-15}$ $\funits$~  and in the hard 
        band at a flux of $1 \times 10^{-14}$ $\funits$. Note that the
        soft band normalization takes into account the different Galactic 
        hydrogen column density.
\end{itemize}
\end{table}
\noindent
The counts agree with recent studies of these
fields \citep{Gia01, To01, Ro02, Br01, Wa04}. The slope values match
the published ones within the errors. The normalizations are also
consistent within the errors, considering that we used different, more
recent calibrations and a different spectral slope ($\Gamma$ = 1.7) to
derive the conversion factors (see Appendix~\ref{appendix:Sec2}).
Most of the authors quoted above use a flatter slope with $\Gamma$ =
1.4. The average difference between the conversion factors obtained
considering an absorbed power law spectrum with $\Gamma$ = 1.7 and
$\Gamma$~=~1.4 are

\noindent
\centerline{$K_{soft}(1.4) \sim  K_{soft}(1.7) - 3\% K_{soft}(1.7)$}
\noindent
\centerline{$K_{hard}(1.4) \sim  K_{hard}(1.7) + 12\% K_{hard}(1.7)$~.}

\smallskip\noindent
Finally the four reference fields were combined in a unique big 
field of $\sim 0.31~\rm deg^{2}$. A corresponding sky coverage was 
also constructed.  The 378 plus 280 sources, detected in the soft and  
hard band respectively, have been used to derive both
the differential and integral $\log~N{-}\log~S$ for the {\em unified 
reference field}. The data were best-fitted using the MLM. The best-fit 
integral source count parameters are listed in Table~\ref{tab3}. 

\begin{figure}
\begin{center}
\includegraphics[width=14cm, height=14cm, bb=0 -30 574 544]{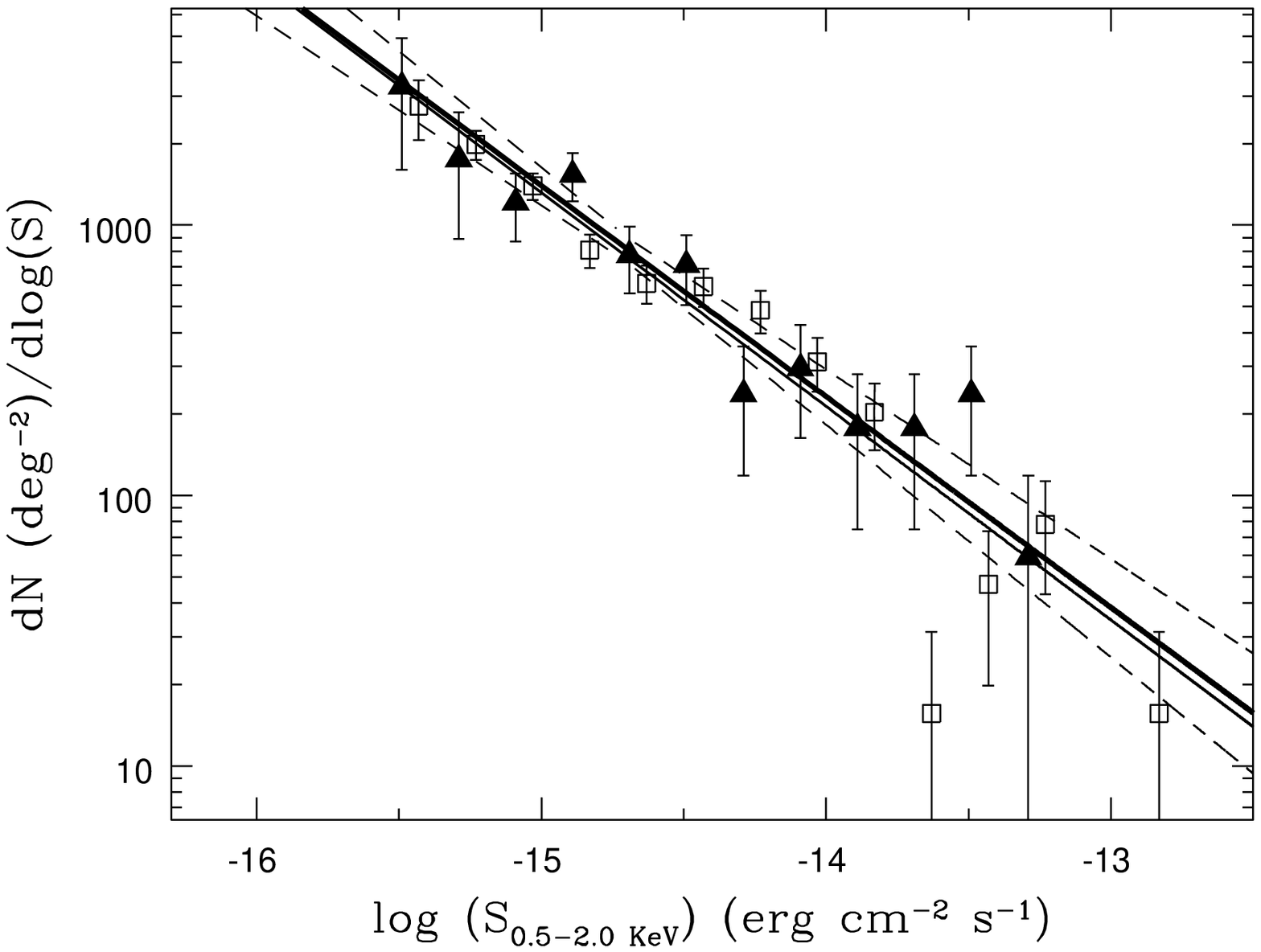}
\end{center}
\vspace{-5.5cm}
\begin{center}
\includegraphics[width=14cm, height=14cm, bb=0 -30 574 544]{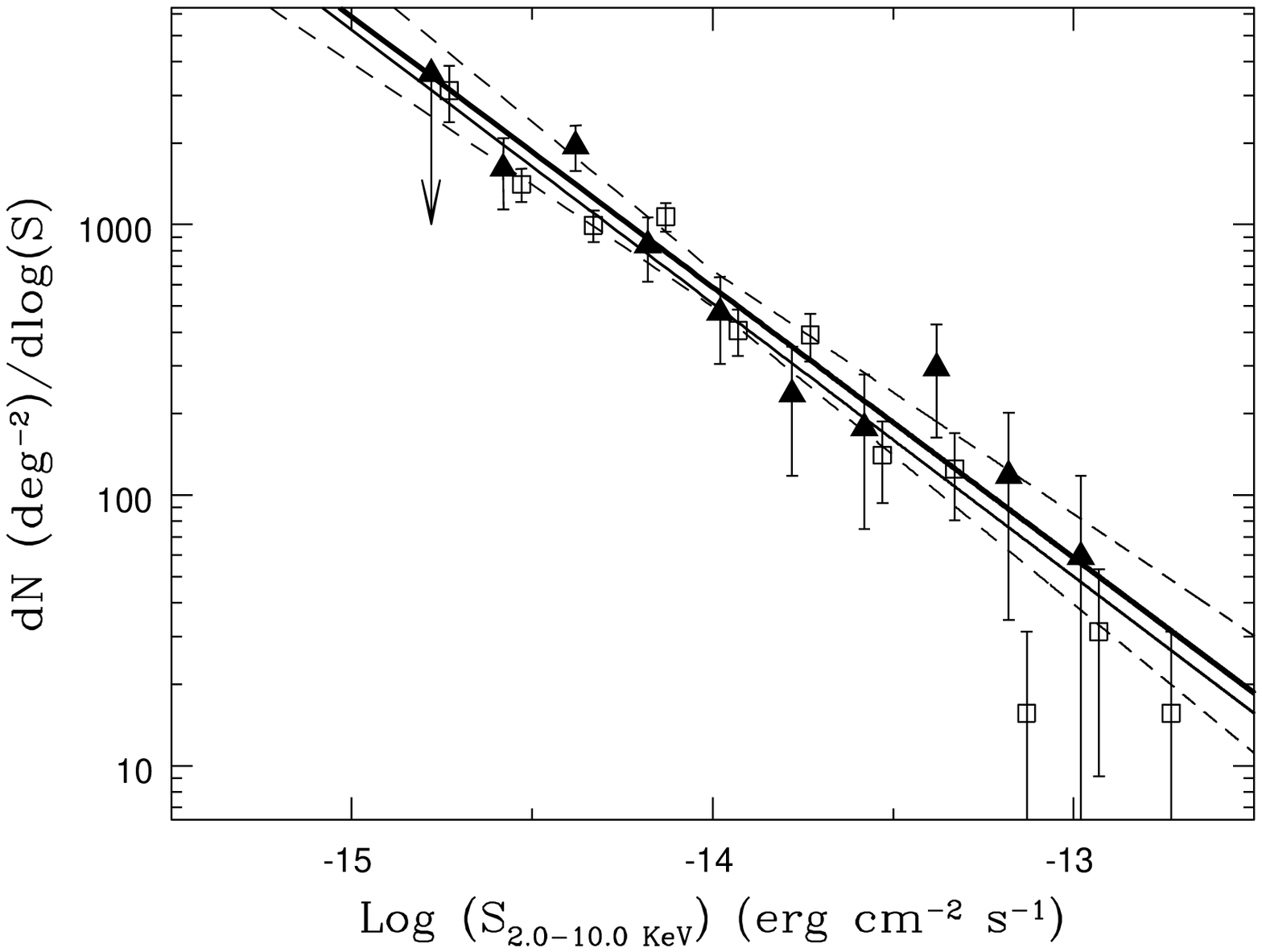}
\end{center}
\vspace{-5.0cm}
\caption{Differential $\log~N{-}\log~S$s. 
The differential $\log~N{-}\log~S$ of the X--ray point sources 
detected in areas covered by the clusters are indicated by solid triangles
(soft band in the top panel  and hard band in the bottom panel).
Open squares indicate the differential $\log~N{-}\log~S$ of the X--ray point 
sources detected in the field obtained by combining the four deep 
($\tau \sim 100 ks$) fields without clusters (Bootes, CDFS, HDFN, and 
Groth Strip). The uncertainty on each point corresponds to a 1 $\sigma$ 
confidence level. No cluster region sources are observed in the hard faintest 
flux bin. The arrow indicates an upper limit corresponding to three sources.
According to Poisson statistics, we have a 5\%  probability of observing
zero sources when three are expected.
The thick (thin) solid lines represent the best-fit power law of the 
$\log~N{-}\log~S$ for the cluster region (field) sources obtained with the  
MLM, while the dashed lines correspond to the 1 $\sigma$ uncertainties on the 
fit for the cluster region sources. The uncertainties on the field source fit 
are smaller and have been omitted for clarity.} 
\label{fig1}
\end{figure}

\noindent 
Hereafter the {\em unified reference field} is
used for comparison with the cluster fields, but the four single
reference fields give a measure of the parameter dispersion.  The
differential source counts in bins of $\Delta \log S=$ 0.2 for the
{\em unified reference field} are indicated by open squares in
Fig.~\ref{fig1}.  The errors correspond to 1 $\sigma$ confidence level
obtained as in Eq.~\ref{eqn5}. The thin solid lines represent the power
law corresponding to the best-fit parameters obtained with the MLM. 

\begin{figure}
\begin{center}
\includegraphics[width=14cm, height=14cm, bb=0 0 574 574]{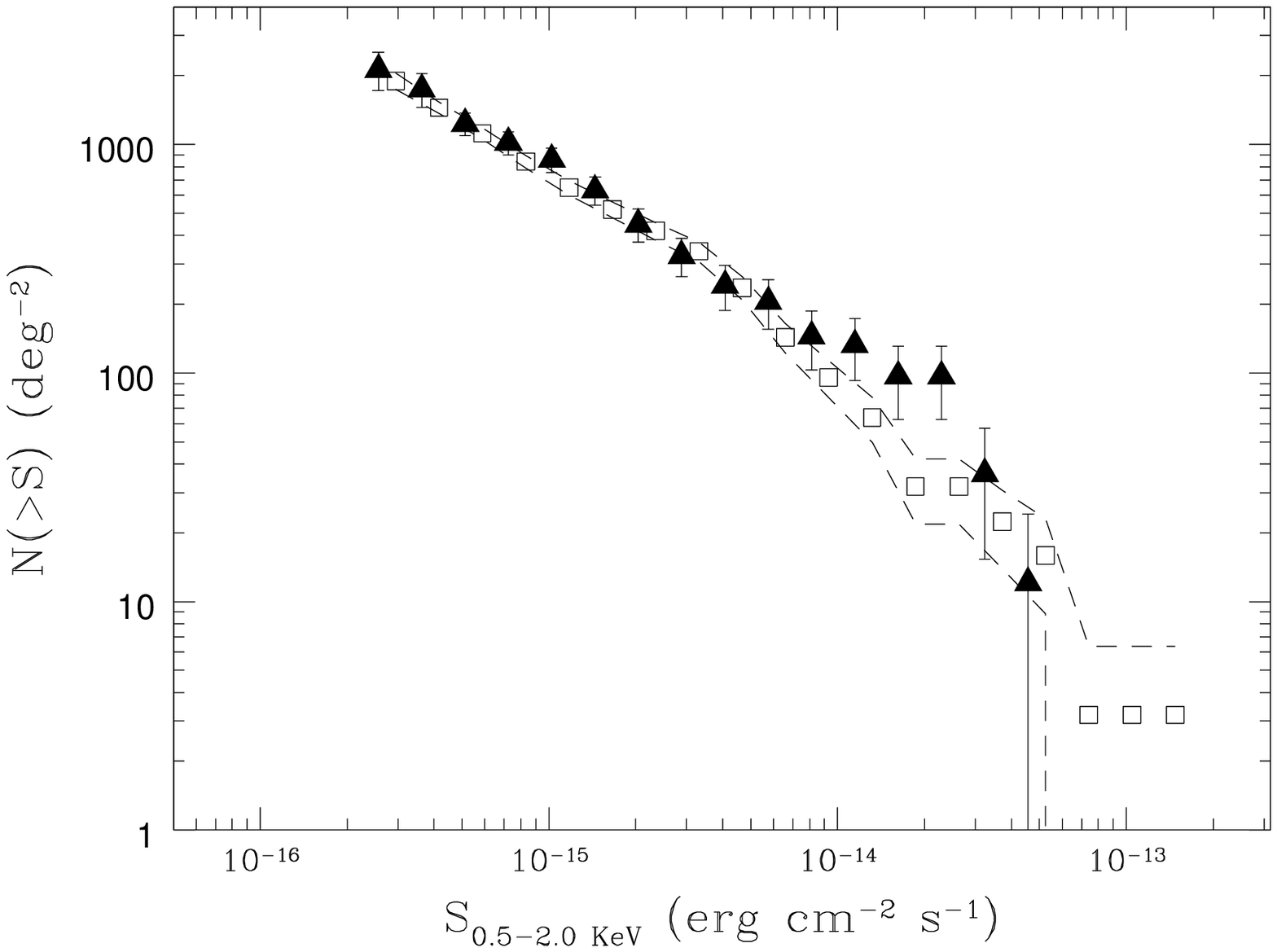}
\end{center}
\vspace{-5cm}
\begin{center}
\includegraphics[width=14cm, height=14cm, bb=0 0 574 574]{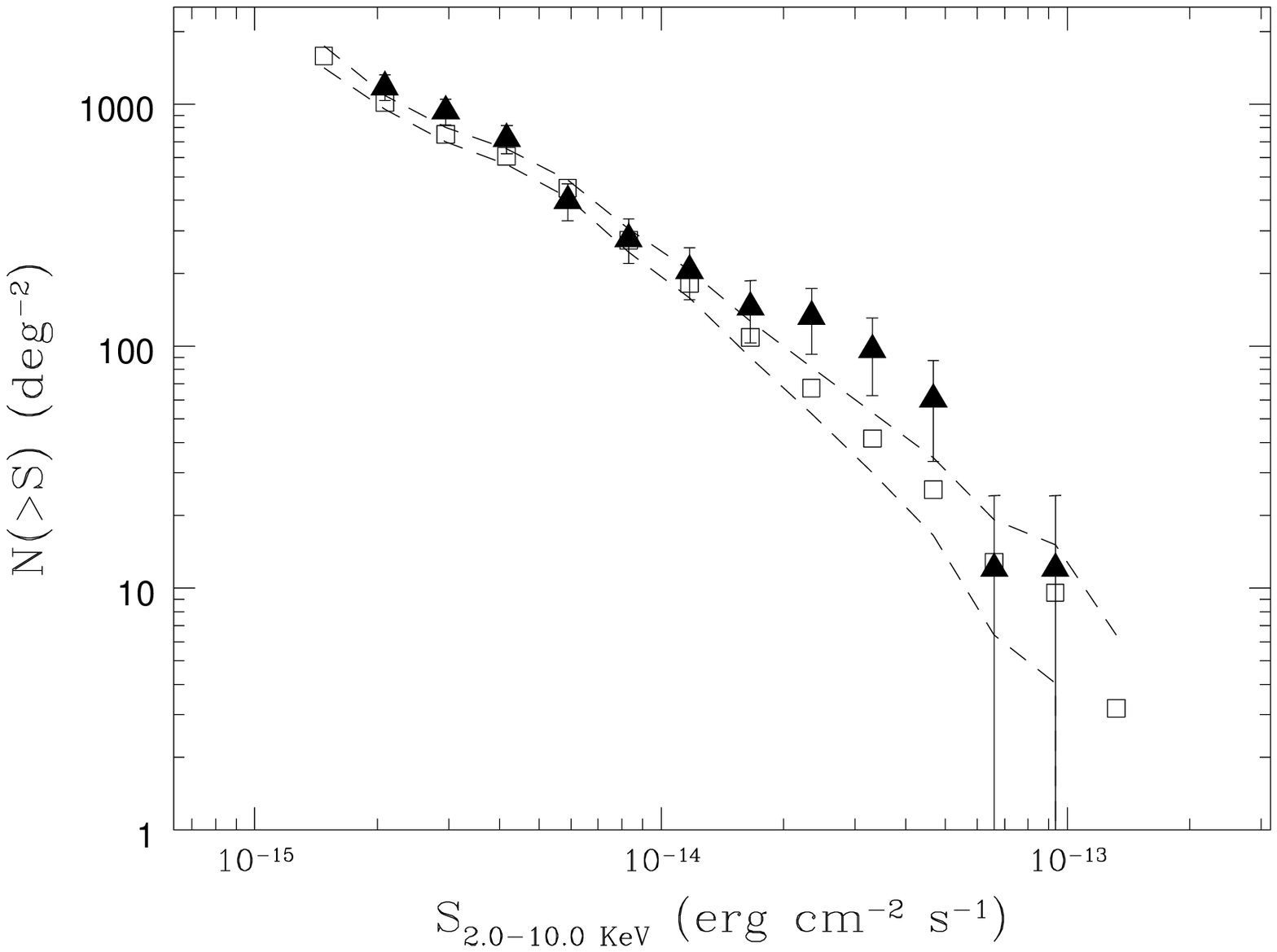}
\end{center}
\vspace{-4.5cm}
\caption{Integral $\log~N{-}\log~S$s.The integral $\log~N{-}\log~S$ for 
the sources detected in the regions occupied by clusters are indicated by 
filled triangles in both the soft (top panel) and hard (bottom panel) 
energy band. Open squares indicate the integral $\log~N{-}\log~S$ for the 
sources detected in the {\it unified reference field} obtained by combining 
the four $\sim 100 ks$ long fields without visible clusters (Bootes, CDFS, 
HDFN, and Groth Strip). The cumulative source counts  are plotted using a step 
$\Delta$\,log S $=$ 0.15. Uncertainties on each  $\log~N{-}\log~S$  point
correspond to a 1 $\sigma$ confidence level obtained as described in the 
text. Dashed lines represent 1 $\sigma$ confidence level for the field 
sources counts.} 
\label{fig2}
\end{figure}

\smallskip\noindent
The cumulative source counts of the {\em unified reference field} 
are shown in Fig.~\ref{fig2} (open squares). The dashed lines represent
1 $\sigma$ uncertainties on the source counts.

\section{Are X--ray point sources associated with the clusters?}
\subsection{Cluster counts vs field counts}
\label{disc}

\begin{figure}
\begin{center}
\includegraphics[width=8cm, height=8cm, bb=50 150 624 724]{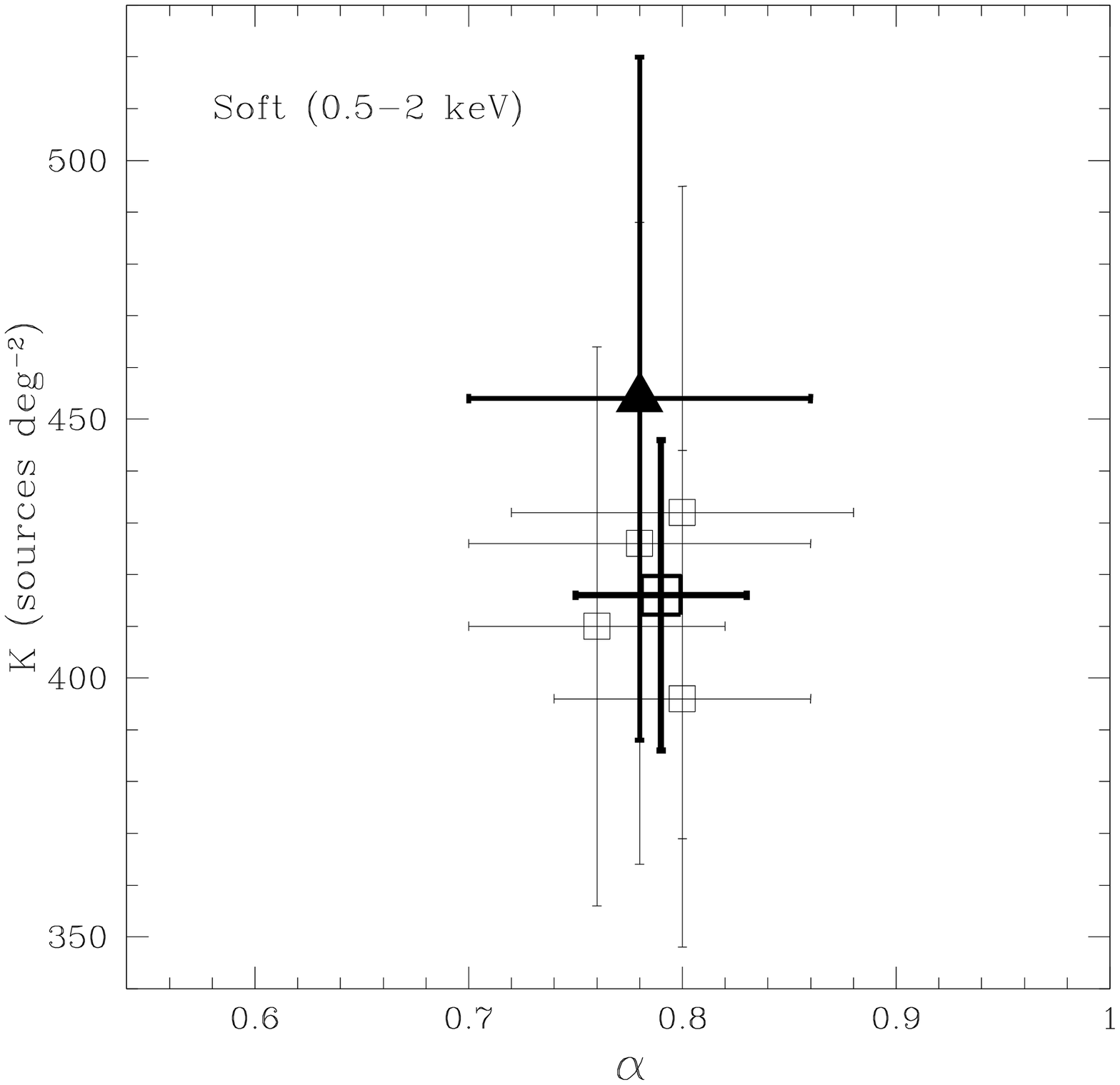}
\includegraphics[width=8cm, height=8cm, bb=0 150 574 724]{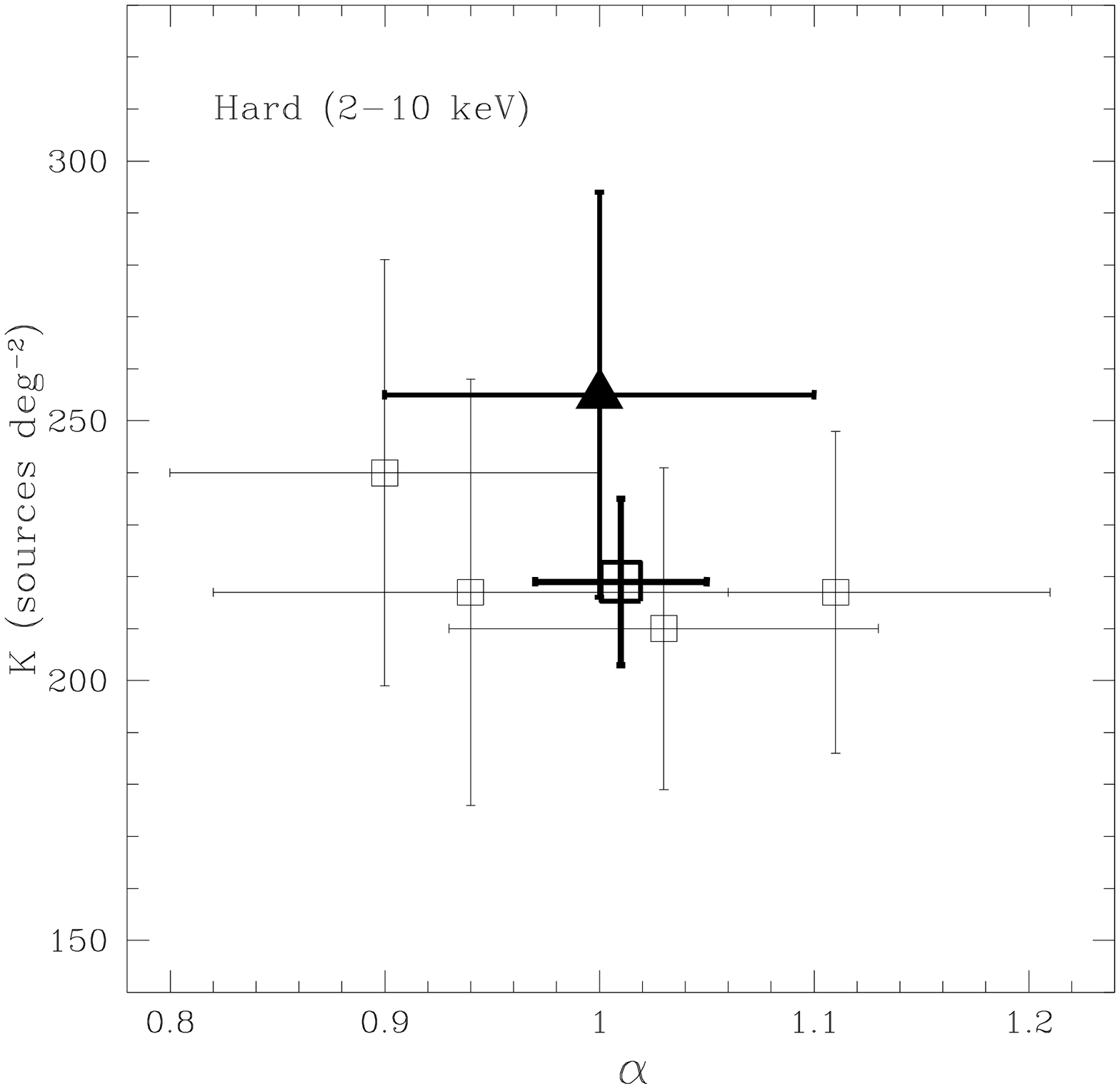}
\end{center}
\vspace{-0.5cm}
\caption{
Integral $\log~N{-}\log~S$ normalization vs slope for both cluster region 
and field sources obtained using the MLM. Solid triangles 
indicate the best-fit results for the cluster region sources (soft band 
to the left and hard band to the right), while the thick open squares 
indicate the {\it unified reference field} results. The thin open squares 
indicate the best-fit results for each of the four reference fields. 
Uncertainties correspond to a 1 $\sigma$ confidence level on the fit parameters.}
\label{fig3}
\end{figure}

Figure~\ref{fig3} illustrates the best-fit normalization and slope of
the integral $\log~N{-}\log~S$s, for both the cluster and the 
reference field sources. The parameters of the fits, given in
Table~\ref{tab3}, were obtained by fitting one single power law
over the entire flux range.  All values are consistent within
1 $\sigma$ confidence level in both bands, showing no significant
difference between the counts in the cluster areas and in the control
fields. However, an inspection of the top panel of
Fig.~\ref{fig2} shows that the soft integral cluster counts
exhibit a small excess at the bright end around $\sim2\times10^{-14}$
$\funits$. A similar excess is present in the hard
integral number counts (Fig.~\ref{fig2}, bottom) around $\sim 3.0
\times 10^{-14}$ $\funits$.
%
\noindent
The significance of these apparent excesses was evaluated by computing
the number of field point sources expected in the cluster regions.
Since the field source counts in both bands show a steeper slope at
the bright end compared to the faint end, as also found
by several authors \citep{Ha98,Mu00,To01,Ro02}, instead of using
the single power-law best fit, we considered it more appropriate
computing the expected number of field sources by directly normalizing 
the number of the observed point sources in the control fields to the
area covered by the clusters. We would expect 5$\pm$1 field sources
above a flux of $1.25 \times 10^{-14}$ $\funits$ for the soft band and
the same number above $2.5 \times 10^{-14}$ $\funits$ for the hard
band. We instead find 11 sources in the cluster region in each
band. The significance of the difference is at the 1.7 $\sigma$ level
in both cases. Note that 9 sources out of the 11 are common to both
bands.  {\bb If this small excess is real, we are facing a cluster
population of X--ray sources with X--ray luminosities ranging from
$\approx 3 \times 10^{42}$ to $10^{44}$ erg $s^{-1}$ in the soft band
and from $\approx 10^{43}$ to $10^{44}$ erg $s^{-1}$ in the hard
band}.

\smallskip\noindent
Next we examined whether there is any dependence of the number 
counts on cluster redshift. For this purpose the point source sample was 
divided  into two subsamples: the ``high--z'' subsample corresponding to the 
sources detected in the regions covered by clusters with redshift z $>$ 0.7 
and the ``low--z'' subsample corresponding to the sources detected in the  
regions covered by clusters with redshift z $<$ 0.7.  {\bb We performed an 
analysis on these two subsamples similar to the one performed
on the whole sample. No significant difference between the two subsamples 
is evident in the soft band. In the hard band there is instead a slight
indication of a possible excess at bright fluxes in the ``high-z'' subsample 
with respect to the ``low-z'' one.  The statistics are too poor to draw any
conclusion}.

\smallskip\noindent 
{\bb Finally it is interesting to compare our
estimate of the ``cluster X--ray population'' with the results by
\cite{Ma06} for a sample of nearby clusters.  Under the assumption
that our clusters are similar to the Martini clusters, we used their
spectroscopically confirmed cluster X--ray sources to derive a
``Cluster X--ray Luminosity Function'' (Cl-XLF) and predict how many
cluster X--ray sources we should expect in our sample.  In their five
clusters with z $>$ 0.15, \cite{Ma06} list seven objects within 1~Mpc
from the cluster center with a soft $L_x > 10^{42}$ erg
$s^{-1}$, while no object is present in the three closer clusters
with redshifts of 0.07, 0.06, and 0.059. If we build the Cl-XLF from the
five z$>$0.15 clusters by properly accounting for our luminosity
limits, we expect five cluster X--ray sources, four of which are 
within 0.5 Mpc and one between 0.5 and 1 Mpc from the cluster
center\footnote{Since the search radius $\rm R_{ext}$ does not extend
up to 1.0 Mpc for all the clusters, the number of sources expected in
each annulus was normalized by the area actually examined.}.  These
expected numbers agree with the  numbers of sources in clusters previously
estimated from our own data. Had we used the data from
all the Martini eight clusters, the Cl-XLF would have had a lower amplitude,
and the predicted number of cluster X--ray sources in our sample would have
dropped to $\le~4$.  A similar analysis cannot be performed in the hard
band due to the presence of only two X--ray sources in \cite{Ma06} work.}

\subsection{Radial distribution of the point sources}

To further investigate the reality of the excess at high flux, we  
examined the radial distribution of the brightest sources {\bb in 
both bands} as a function of the projected linear distance from the 
cluster center. Given the small number of sources, only two 0.5 
Mpc wide annuli were used. From the control field source counts, {\bb 
normalizing to the areas actually examined}, we computed the expected 
number of field sources in the two cluster annuli. The results are 
listed in Table~\ref{tab4} (Columns (2) and (5)). In each band 5 field 
sources are expected in the area covered by the clusters 
(see Sect.~\ref{disc}). Two field sources are expected in the inner 
annulus against 7 observed, and 3 field sources are expected in the 
external annulus against 4 observed. There is a 2 $\sigma$ excess 
within 0.5 Mpc from the cluster center, thereby confirming the small excess 
of sources found in Sect.~\ref{disc} and providing additional support 
for the possible existence of X--ray sources belonging to the cluster 
(on average one every three clusters).
\begin{table}[htb]
\begin{center}
\caption{Radial source distribution for the brightest sources}
\scriptsize
\small
\begin{tabular}{lcccccccccc}
\hline 
\hline
R &~~~~&$\rm N_{0.5-2.0}$ & $\rm \Sigma_{0.5-2.0}^{Cl}$ & $\rm \Sigma^{field}_{0.5-2.0}$ &~~~~~~~~~&$\rm N_{2.0-10.0}$ & $\rm \Sigma_{2.0-10.0}^{Cl}$ & $\rm \Sigma ^{field}_{2.0-10.0}$ \\
(1) & & (2)& (3)& (4)& & (5)& (6)& (7) \\
\hline
\hline
\\
0--0.5 &  & 7 (2) & 0.50~$\pm$~0.19 & 0.15~$\pm$~0.03 & & 7 (2) & 0.50~$\pm$~0.19 & 0.13~$\pm$~0.03\\
\hline
\\
0.5--1.0 & & 4 (3) & 0.18~$\pm$~0.09 & 0.15~$\pm$~0.03 & & 4 (3) & 0.18~$\pm$~0.09 & 0.13~$\pm$~0.03\\
\hline
\label{tab4}
\end{tabular}
\end{center}
\vspace{-0.5cm}
\begin{itemize}
\item [--] Column 1: Size of the annulus in Mpc 
\item [--] Column 2: Number of sources detected in the 
    annulus with an unabsorbed flux brighter than $1.25 \times 10^{-14}$ 
    $\funits$ ~in the soft energy band.  In parentheses the expected number 
    of field sources over the cluster regions is given
\item [--] Column 3: Soft X--ray surface density for the cluster region 
                     sources in Mpc$^{-2}$
\item [--] Column 4: Estimated soft X--ray surface density for the control 
                     field sources in Mpc$^{-2}$
\item [--] Column 5: Number of sources detected in the circular
    annulus with a flux brighter than $2.5 \times 10^{-14}$ 
    $\funits$ ~in the hard energy band. In parentheses the expected number
    of field sources over the cluster regions is given
\item [--] Column 6: Hard X--ray surface density for the cluster 
           regions sources in Mpc$^{-2}$
\item [--] Column 7: Estimated hard X--ray surface density for the control
                    field sources in Mpc$^{-2}$
\end{itemize}
\end{table}

\medskip\noindent 
{\bb Next we constructed the radial profile of the surface density, 
$\Sigma^{cl}$, within 1 Mpc (Columns (3) and  (6) of 
Table~\ref{tab4}), assuming the brightest sources to be at the cluster 
redshift and taking the areas effectively surveyed into account.
The field source number density per Mpc$^{2}$, $\Sigma^{field}$
(Columns (4) and (7)), was then derived assuming the same redshift 
distribution as in our cluster sample.}

\medskip\noindent 
The radial profile of the soft source surface
density, $\Sigma^{cl}$, was compared with the distribution
of \cite{Ru05} in a subsample of the MACS clusters.  {\bb The MACS
survey was built to find the ``most'' massive clusters ever
\citep[see][]{Eb01}.  The median MACS X--ray luminosity\footnote{The
MACS median luminosity was estimated using Fig. 10 in Ebeling et
al. (2001) after converting their 0.1--2.4 keV {\em ROSAT} luminosity into
bolometric luminosity with their same assumptions on the temperature,
and taking into account the different cosmologies assumed. } is 2.5
times our median luminosity with only 3 (out of 18) of our clusters
more luminous than the MACS median luminosity.} The flux limit we
adopted in the soft band is the same as the one used by them 
($1.25\times 10^{-14}$ $\funits$). Within 0.5 Mpc, where the excess is
found, the two densities are in very good agreement {\bb (0.50 sources
per Mpc$^{2}$ from Table \ref{tab4} vs 0.53 sources per Mpc$^{2}$ from
their Fig. 2). Within 0.5 Mpc the excess of \cite{Ru05} is at 
$\approx$ 4 $\sigma$ with respect to their point-source density at the 
cluster field edges (4--7 Mpc) that they assume to be the background 
point-source density\footnote{The significance of 8.0 $\sigma$ claimed 
by \cite{Ru05} is actually for the area within 3.5 Mpc from the 
cluster centers.}. Between 0.5 and 1 Mpc, the two profiles are consistent 
within the errors. We stress that the field surface density obtained 
from our control fields agree very well with the one derived 
by \cite{Ru05} from their cluster field edges. } 

\medskip\noindent
In order to increase the statistics, we extended our radial analysis to 
fainter sources. We used 31 sources with a  
soft unabsorbed flux brighter than $2.7 \times 10^{-15}$ $\funits$ and 25 
sources with a hard flux brighter than $0.8 \times 10^{-14}$ $\funits$
{\bb since  all  clusters are complete down to these flux limits.
A small source excess is found within 0.5 Mpc from the clusters
center, which however does not improve significantly the previous result,
indicating that cluster X--ray sources are confined to the highest fluxes.} 

\subsection{Optical counterparts}

We checked the literature for any optical counterpart associated with
the 13 X--ray sources with a soft flux brighter than $1.25 \times
10^{-14}$ $\funits$ and/or a hard flux brighter than $2.5 \times
10^{-14}$ $\funits$.  {\bb The results of this search are reported in
Table~\ref{tab5}. No optical data are available for four sources. Two
sources are identified with cluster members with spectroscopic
redshifts. Three are identified with optically faint objects with no
 redshift measurement. Finally three sources are background objects
having a significantly higher redshift than those of the corresponding
clusters. A fourth one may also be a background object on the basis of
a photometric redshift. Clearly more optical work is needed in order
to know the cluster membership of these objects.}
\begin{table}[htb]
\begin{center}
\caption{Optical parameters for the 13 X--ray brightest sources}
\begin{tabular}{lccccclc}
\\
\hline
\hline
Cluster name& \# & RA & DEC & $\Delta$RA(x-o) & $\Delta$Dec(x-o) &~~~~z & Notes\\ 
            &    & $~~hh ~mm ~ss.s$ & ~~\deg ~~~~\arcmin ~~~~\arcsec  & sec & \arcsec & \\
(1) & (2) & (3) & (4) & (5) & (6) &~~~(7) & (8)\\
\hline
\hline
Abell\,2125          &  5& 15 40 56.4&    +66  16  28.7 & +0.0& -0.3 & 1.012 & a\\ 
\\   	     	     				        
ZW\,CL\,0024$+$1652  &  2& 00 26 31.1&    +17  10  17.3 & +0.0& +0.7 & 0.400*& b\\
\\	     	     				        		   
MS\,1621$+$2640      &  4& 16 23 43.7&    +26  32  44.7 & -0.1& -0.3 & ----- & c\\
\\		      
CL\,1641$+$4001      &  2& 16 41 50.3&    +40  01  45.7 & --- & ---  & ----- & -\\ 
   	    	     &  3& 16 41 54.2&    +40  00  32.6 & +0.0& +0.6 & 1.003 & d\\  	       
\\		     	     						       
V\,1524$+$0957       &  7& 15 24 43.4&    +09  55  36.0 & --- & ---  & ----- & -\\  
\\		     	     							      
V\,1121$+$2327       &  3& 11 20 54.0&    +23  27  04.9 & --- & ---  & ----- & -\\    	     
\\		     
V\,1221$+$4918       &  3& 12 21 20.1&    +49  18  44.0 & --- & ---  & ----- & -\\    	  
\\		     	     							      
MS\,1137$+$6625      &  4& 11 40 31.2&    +66  08  58.2 & +0.0& +0.0 & 1.269 & e\\
\\		     	     							      
RDCSJ\,1350$+$6007   &  8& 13 50 57.7&    +60  08  13.7 & +0.1& -2.2 & -----  & c\\
\\		     	     							       
RXJ\,1716$+$6708     &  1& 17 16 36.9&    +67  08  30.0 & +0.0& +1.0 & 0.795*& f\\
                     &  4& 17 16 51.7&    +67  08  54.8 & -0.0& -1.1 & ----- & g\\
\\			     				  
MS\,1054$-$0321      &  3& 10 56 58.8&    -03  38  51.2 & -0.1& -0.2 & 1.200$^{ph}$ & h\\  
\hline
\label{tab5}
\end{tabular}
\end{center}
\vspace{-0.5cm}
\begin{itemize}
\item [--] Column 1: Cluster name 
\item [--] Column 2: Source identification number as in Table~\ref{tab2}
\item [--] Column 3-4: X--ray source position; Right Ascension and 
           Declination  (Equatorial J2000, HH MM SS.S, +DD MM SS)
           as in Table~\ref{tab2}
\item [--] Column 5-6: Offset between the position of the X--ray source
           and that of the optical counterpart
           ($\Delta$\,RA in seconds of time and $\Delta$\,DEC in arcsec) 
\item [--] Column 7: Spectroscopic or photometric (indicated by $^{ph}$) 
           redshift. Asterisk indicates cluster membership 
\item [--] Column 8: Literature source: a) \cite{Mi04},
           b) \cite{Cz01}, c) \cite{Ec05}, d) from the Sloan Digital Sky 
           Survey, e) \cite{Si05}, f) \cite{Gi99}, g) from a Keck image 
           taken by I. Gioia and h) \citep{Fo06}.
\end{itemize}
\end{table}

\section{Comparison with radio sources in X--ray selected clusters}

As mentioned in the introduction, radio galaxies in both nearby and
distant clusters have a centrally peaked distribution. \cite{Br06}
analyzed a sample of VLA radio sources detected in 18 X--ray selected
clusters \citep{Gi01} extracted from the NEP survey with redshift and
luminosity distributions similar to the present sample. They
found a pronounced peak of radio sources within 0.2 Abell radii,
corresponding \citep[in the cosmology adopted in][]{Br06} roughly to
the 0.5 Mpc size of the first bin in Table~\ref{tab4} here.  The radio
source surface density associated to this peak is $\approx$ 10
Mpc$^{-2}$, i.e. $\sim$ 20 times higher than the X--ray point source
surface density found here.  However, the larger radio excess could be
a selection effect due to the better sensitivity of the radio
observations.  
\cite{Fa04} discuss a correlation between the 3--9 keV luminosity and 
the 5 GHz core radio luminosity ($\nu \rm L_\nu$) in radio-loud AGN. 
To estimate the X--ray luminosity expected for the \cite{Br06} radio 
galaxies, we first converted the total radio luminosity to 
the core radio luminosity using the correlation between the core and 
radio powers published by \cite{Giov88}. From the resulting core radio 
luminosity and the Falcke correlation, we find that the estimated X--ray 
luminosity of the NEP radio sources is about two orders of magnitude 
lower than the X--ray luminosity reached by the present {\em Chandra} 
observations. Thus the X--ray instruments do not seem to have the 
sensitivity required to detect such faint X--ray counterparts of the 
radio-loud AGN.  
Similar arguments apply if the weaker NEP radio sources are star-forming
galaxies due to the tight linear relations between the X--ray, radio,
and far infrared luminosities found, among others, by \cite{Ra03}.

\smallskip\noindent 
On the other hand, the dozen X--ray sources at the bright end of 
the $\log~N{-}\log~S$ have a high enough X--ray
luminosity to statistically expect that radio emission from some of
them would be detectable in surveys like the FIRST \citep{Bec95} or
the NVSS (NRAO VLA Sky Survey, \citealt{con98}).  Indeed two such
X--ray sources have associated radio emission.  Source \# 3 in
CL\,1641$+$4001 has a FIRST counterpart at the position
$16^{h}41^{m}54.24^{s}$, $+40^{\circ}00^{'}32.0^{''}$ with a flux
S$_{1.4 GHz}$ = 5.06 mJy. Source \# 1 in RXJ\,1716.4$+$6708 has a NVSS
counterpart at the position $17^{h}16^{m}37.14^{s},
67^{\circ}08^{'}28.8^{''}$ with a flux S$_{1.4 GHz}$ = 332.0 mJy (4C
+67.26). Given the low statistics no conclusion is drawn here.

\smallskip\noindent 
For completeness the coordinates of all sources
listed in Table~\ref{tab2} were cross-correlated with those of the
FIRST (or NVSS catalogs when no FIRST data are available). Only one
additional coincidence was found for source \# 2 in
ZW\,CL\,1454.8$+$2233. The corresponding FIRST radio source 
at $14^{h}57^{m}10.82^{s}$, $+22^{\circ}18^{'}44.9^{''}$ 
has a flux of S$_{1.4 GHz}$ = 4.87 mJy.

\section{Summary and conclusions}

In this paper we have presented an analysis of the X--ray point
sources detected in the inner 1~Mpc region of 18 high-z (0.25 $<$ z
$<$1.01) galaxy clusters retrieved from the {\em Chandra} archive.
Unlike most of the previous studies that analyzed the whole
{\it Chandra} field around the clusters, we considered only the point sources
embedded in the cluster emission, i.e. belonging to the clusters or in
projection. {\bb We find a small excess} for the cluster sources at
fluxes brighter than $1.25 \times 10^{-14}$ $\funits$ in the soft
energy band and brighter than $2.5 \times 10^{-14}$ $\funits$ in the
hard energy band. The significance of the excess is at the 1.7
$\sigma$ level in each band (Sect.~\ref{disc}).  To further
investigate the reality of the X--ray point source excess, we examined
the source radial distribution as a function of the projected linear
distance from the cluster center. A 2 $\sigma$ excess was found within
0.5 Mpc providing additional support for the existence of X--ray
sources belonging to the cluster.  The excess is given by $\approx$ 6
out of 11 sources in the luminosity range ~3$\times$10$^{42}$ --
10$^{44}$ erg s$^{-1}$ in the soft and 10$^{43}$ -- 10$^{44}$ erg
s$^{-1}$ in the hard energy band. These results agree with those of 
previous studies that have detected excesses in cluster fields.

\medskip\noindent 
The galaxy clusters presented here are in the bolometric luminosity range 
10$^{44}$ -- $\sim 5\times10^{45}$ erg s$^{-1}$, with a median value of
0.9$\times10^{45}$ erg s$^{-1}$ (only three clusters have a luminosity 
higher than $2\times10^{45}$ erg s$^{-1}$).  \cite{Ru05} examined a sample 
of 51 clusters in a similar redshift range (0.3$<$ z $<$ 0.7), but using
more massive clusters than ours.  They found a soft excess within 1 Mpc
similar to ours. \cite{Je06} analyzed six groups of galaxies in the
redshift range 0.2 $<$ z $<$ 0.6 and obtained a 2 $\sigma$ overdensity 
result for these lower-mass systems. Therefore our results provide
further evidence of the presence of a population of AGN in systems
with very different masses. While \cite{Ru05} did not analyze sources
in the hard band, \cite{Je06} found no significant excess in the
number of hard sources. Thus our study provides for the first time  
some evidence of source overdensity within 1 Mpc in both energy
bands. {\bb In addition, there is a slight indication that the hard band 
excess increases with redshift, even though the small statistics do not
allow us to draw any conclusion. We note, however, that this effect goes
in the same direction as the apparent correlation between the
amplitude of the overdensity and the cluster redshift found for the
first time by \cite{Ca05} \citep[see also][]{Ma06}.}

\smallskip\noindent 
Radio galaxies in both nearby and distant clusters
have a centrally peaked distribution with surface density within 0.5
Mpc on the order of 10 Mpc$^{-2}$.  A recent work by \cite{Br06} on a
sample of distant X--ray selected clusters (0.3 $<$ z $<$ 0.8) finds
an excess in the radio surface density within 0.5 Mpc from the cluster
center that is much higher (by a factor $\sim$ 20) than the present
X--ray source overdensity.  Even if an excess of sources is present at
both wavelengths, the much smaller amplitude of the X--ray overdensity
could be explained by the better sensitivity of the radio
observations.  Higher sensitivity (and resolution) X--ray telescopes
could provide more information on the nature of the population of AGN
and/or star-forming galaxies in high-z clusters, which is responsible
for the more pronounced excess in the radio domain.

\begin{acknowledgements}
We acknowledge stimulating discussions with Paolo Tozzi, Anna Wolter and 
Mauro Dadina. An anonymous referee gave substantial comments which helped 
to improve this paper. Partial financial 
support for this work came from the Italian Space Agency ASI (Agenzia 
Spaziale Italiana) through grant ASI-INAF I/023/05/0.
\end{acknowledgements}

\clearpage
\appendix

\section {Data reduction}
\label{appendix:Sec1}
We retrieved the level$=$1  event files from the archive and applied the 
standard processing. The CIAO  tool \textit{acis\_process\_events} was 
used to apply the correction for charge transfer inefficiency 
\citep[CTI][]{To00, Gr05} and recomputation of event grades. 
To compute calibrated photon energies, \textit{acis\_process\_events} 
was also used to update the Advanced CCD Imaging Spectrometer 
({\em ACIS}) gain maps with the latest version provided within CALDB (ver. 
3.0.3) and to correct for its time dependence (T\_GAIN correction). 
The CTI and T\_GAIN corrections were applied for those chips and period 
observations for which they were available.
Most of the observations were telemetered in VFAINT mode, which provides 
a better rejection of the particle-induced background. For these 
observations we ran \textit{acis\_process\_events} with the option 
`check\_vf\_pha=yes', to flag probable background events. 
After running \textit{acis\_process\_events}, the events were filtered to 
include only the standard event grades 0, 2, 3, 4, and 6 and status 
bits set to~0. In this way we removed photons detected in bad CCD columns 
and bad pixels, `problem' events such as cosmic ray afterglows, and also those 
with bad {\em ASCA} grades (1, 5, and 7).

\smallskip\noindent
The final step was to examine the background light curves during each 
observation to detect and remove the flaring episodes. After excluding 
the point sources and cluster emission from the event file, the script 
\textit{lc\_clean} was used to extract and bin the light curve 
and to calculate the average count rate. 
The flare detection was performed following the recommendations
on the energy band and the bin size given in the Markevitch COOKBOOK 
\footnote{See {\em http://cxc.harvard.edu/contrib/maxim/acisbg/data/README} 
and \\ {\em http://cxc.harvard.edu/contrib/maxim/acisbg/COOKBOOK}}.
We excluded those time periods when the count rate was not 
within 20\% of the quiescent rate.
The final  count rate in the source-free regions of each observation 
was compared with the background value tabulated by Markevitch.
We obtained values consistent within less than 10\%.
Compared to the {\em ACIS} front-side illuminated (FI) chips the back-side  
illuminated (BI) chips (S1 and S3) have a higher sensitivity at low energies. 
However, this low-energy sensitivity makes the chip  more sensitive 
to particle events, which results in more frequent background flares 
than in the FI chips \citep{Pu00,Mar01,Mar03}. 
When the source lay in the S3 chip, the S1 chip was accurately 
examined to exclude completely the flare-affected period.

\smallskip\noindent 
The released calibration underestimates the effective area of 
the {\em Chandra} mirror by 10\% just above the Ir M edge, probably 
because the mirror surface is contaminated by a thin hydrocarbon 
layer \citep{Mr03}. To correct the effective area, a 
``positive absorption edge'' described by \cite{Vi05} was used in the analysis 
of the spectra of point sources with the X--Ray Spectral 
Fitting Package (\textit{XSPEC}). 

\section{Computation of Source Parameters} 
\label{appendix:Sec2} 

The net counts, C, calculated as the sum of all counts in the ``source cell" 
subtracted by the sum of the background counts, B, were used to  estimate the 
source flux. Since the majority of the detected sources have poor statistics
(less than 50 total net counts in the soft or the hard band), an estimate of 
the source flux through a fit to the data with an absorbed  power law is not 
always possible.  Two separate conversion factors, one for each band, 
were then calculated to derive the flux ($S$) from the observed 
count rate.

\smallskip\noindent 
The net count rate for each source was computed by dividing the net counts 
by the effective exposure time at each source position. For each source,
the effective exposure time is given by the observation exposure time 
(corrected for the flares' time periods) multiplied by the ratio of the 
exposure map, averaged within the extraction region for each 
source to the value of the exposure map at the aimpoint.
The vignetting correction (V) to be applied to the net counts
is given by the ratio of the value of the exposure map at the aimpoint 
($expmap\_aimpoint$) to the value of the exposure map at the source 
position ($expmap\_source$).  The correction is done separately for each 
band using the exposure maps computed at energies of 1.0 keV 
(soft band) and of 4.0 keV (hard band). In this way the source count 
rate corresponds to the count rate that the source would have if it were 
observed at the aimpoint. The soft and hard conversion factors ($K$) from 
counts (cts s$^{-1}$) to X--ray fluxes ($\funits$) were derived
at the aimpoint using the response matrices of the detector 
at this position. Such conversion factors were computed assuming an absorbed 
power-law spectrum with a photon index $\Gamma=$1.7 \citep{Mu84,WE87} and 
assuming for the hydrogen column density, N$_H$, the Galactic value
along the line of sight at the source position (see Table~\ref{tab1}, 
Column 9, and Table~\ref{tab3}, Column 4). 

\smallskip\noindent
The soft and hard X--ray source fluxes were calculated as 
\begin{eqnarray}
\label{eqn9}
S = \frac{C}{\tau} \times \frac{expmap\_aimpoint}{expmap\_source} \times K
  = \frac{C}{\tau} \times  V \times K
\end{eqnarray} 
where $C$ are the net counts, $\tau$ the flare-corrected exposure time,
$V$ the vignetting correction factor, and $K$ is the conversion factor 
from counts s$^{-1}$ to X--ray fluxes in $\funits$~ appropriate for each 
energy band. 
The conversion factors allowed us to convert the 0.5--2.0 keV band count 
rate to the observed X--ray fluxes in the same band and the 2.0--7.0 keV band 
count rate to the observed X--ray flux in the 2.0--10.0 keV band. 

\smallskip\noindent
Flux uncertainties were estimated taking into account  the error
on the net counts ($\sigma_{NC}$) computed as the square root of
the total observed counts in the ``source cell'', i.e. $\sqrt{C + B}$,
and the error on the conversion factor,  $\sigma_{K}$, due to the  assumed 
different power-law models.
The uncertainties on the conversion factor reflect the range of possible 
values for the effective photon index: $\Gamma$ = 1.4--2.0.
The error on the flux then becomes:

\begin{eqnarray}
\sigma_{S}= \sqrt{{\sigma_{NC}}^2 \times {\left( \frac{V \times K}{\tau}\right)}^2 + 
  {\sigma_{K}}^2 \times {\left( \frac{C \times V}{\tau}\right)}^2}
\ = \sqrt{S^2 \times \left( \frac{{\sigma_{NC}}^2}{C^2} + \frac {{\sigma_{K}}^2}{K^2} \right)}~. 
\label{eqn10}
\end{eqnarray}

\smallskip\noindent
The soft observed (i.e. absorbed) X--ray flux used to derived the source 
counts were multiplied by the factor, c$_{N_{H}}$ 
(Table~\ref{tab2}, Column 3) to obtain the Galactic unabsorbed X--ray flux. 
The correction factors were derived taking into account the Galactic hydrogen 
column density, $\rm N_H$, along the line of sight of each cluster 
(Table~\ref{tab1}, Column 9)  and of each void of cluster field 
(Table~\ref{tab3}, Column 4). 

\section{Sky Coverage}
\label{appendix:Sec3}
In Sect.~\ref{cap4} we  derive the source counts (or 
$\log~N{-}\log~S$)  for the point sources detected in the cluster regions. 
An important ingredient for the $\log~N{-}\log~S$ is the determination of 
the sky coverage. The ``sky coverage'' is the area of  sky sensitive down to 
a given  flux limit as a function of the flux density. To estimate the sky 
coverage, $\Omega$, we constructed a flux limit map, indicating 
the flux of the faintest source that would have been included in our 
source list at each position of the {\em ACIS} chip  within the 
region of the cluster. The flux limit map has been constructed to 
account for the following effects: 
\begin{itemize}
\item [--] instrumental effects, such as vignetting, or increase of the 
point spread function size with off-axis angle. The sensitivity of the 
{\em ACIS} detectors varies significantly across the field of view;
\item  [--] background effects. The background considered also includes
  the extended emission from the cluster's hot gas, since we detected 
  sources within the region occupied by the clusters.
\end{itemize}

\smallskip\noindent
Following Eq.~\ref{eqn9} the sensitivity limit ($S_{lim}$ in $\funits$) 
at each detector position was defined as:
\begin{eqnarray}
S_{lim} = \frac{C_{lim}}{\tau} \times  V \times K
\label{eqn11}
\end{eqnarray}
where C$_{lim}$ are the net counts derived as in Eq.~\ref{eqn12}.

\smallskip\noindent
For the chosen signal-to-noise ratio ($S/N=3$; Section~\ref{sample}) 
one obtains from Eq.~\ref{eqn1}

\begin{eqnarray}
C_{lim} = 3 \times (1 + \sqrt{0.75+B_{avg} \times A}) 
\label{eqn12}
\end{eqnarray}

\noindent
where $B_{avg}$ is the average background counts per pixel,  $A$ the area 
expressed as the number of pixels covered by the source cell, and  
$B_{avg} \times  A$ is the local background counts within the source region. 
The estimate of this area $A$ takes into account the degradation 
of the {\em Chandra} PSF,  which increases with the off-axis angle.
To estimate the local background (inclusive of the cluster extended emission) 
and  the  vignetting factor at each position of the chip,
we built both a background and an exposure  map of the same size and 
resolution for both the soft and the hard energy bands. The map of the 
effective background plus the extended cluster emission were obtained as 
follows:
\begin{enumerate}
\item All identified point sources were subtracted from the
soft and hard images. The  resulting ``holes'' were filled with 
pixel values sampled from the Poisson distribution whose mean 
and standard deviation equalled that of the surrounding background 
pixels (using the CIAO tool {\em dmfilth});
\item  These source-free images were binned by 32 $\times $ 32 pixels, 
so that each new pixel covers a linear size of  $\sim$15.7$''$. 
This is a good  compromise between the resolution needed to smooth the 
very local background  variations and  at the same time to sample the 
variations due to the different emission from the cluster;
\end{enumerate}
The exposure maps of the same size and binned by 32 $\times $ 32 pixels 
were constructed at energies of 1.0 keV (soft) and 4.0 keV (hard).
The values corresponding to each image bin ($32 \times 32$ original
pixels) within the region of the cluster allowed the calculation of
the local average background per original pixel and the median
exposure at the position of the bin.  The last step in calculating
$S_{lim}$ is an estimate of $A$.  To do this it is necessary to study
the increasing apparent size of the detected sources as a function of
the off-axis angle.  The size of a detected point source equals, by
definition, the size of the {\it Chandra} PSF, which depends largely
on the source's angular distance from the optical axis and, to some extent,
on source energy. We studied and interpolated the increasing apparent
size of the detected sources as a function of the off-axis angle
separately in the soft and hard energy bands.

\begin{figure}
\includegraphics[width=15cm,height=15cm,bb=0 60 594 654]
                {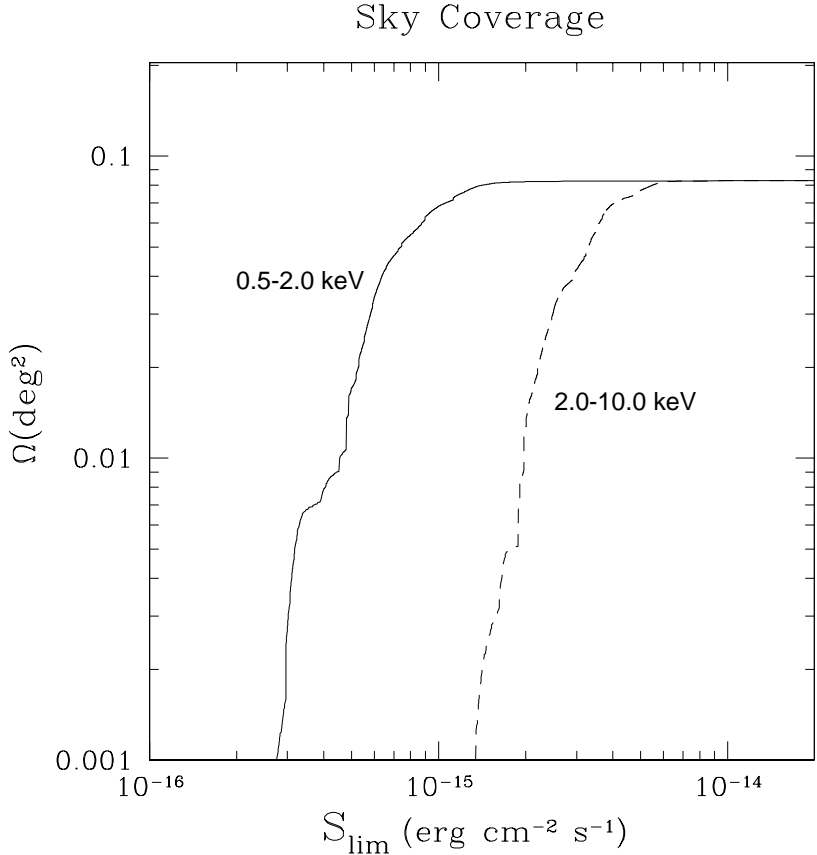}
\vspace{-2.9cm}
\caption{Sky coverage (area covered vs. flux limit) for the soft 
(solid line) and hard (dashed line) energy bands computed for the area 
covered by the 18 clusters of the sample.} 
\label{figC1}
\end{figure}

\smallskip\noindent
For all the clusters we estimated the 
flux limit corresponding to each region. The sky area available at a given 
flux limit is then simply the sum of all the regions whose minimum detectable 
flux is lower than S$_{lim}$. The total area covered by the ``cluster survey" 
is $\sim$ 0.083 deg$^2$. In order to prevent incompleteness effects at the
faint end of the source counts, we considered only those sources for each 
cluster with a flux larger than the flux corresponding to 20\%  of the total 
sky area covered by each cluster. Figure~\ref{figC1} shows the sky coverage for 
the eighteen clusters of our sample computed for the two energy bands. 
The steplike features visible in the sky coverage are given by the
20\% cut described above. Note that the soft fluxes used to derived
the soft-band sky coverage are unabsorbed fluxes, corrected for the 
the Galactic hydrogen column density along the line of sight
of each field.

\end{document}